\documentclass[twocolumn,superscriptaddress,amsfont,amssymb,amsmath, showpacs,balancelastpage, nofootinbib]{revtex4-1}
\usepackage{graphicx,longtable,mathrsfs,color,array}
\usepackage[unicode=true,pdfusetitle,
bookmarks=true,bookmarksnumbered=true,bookmarksopen=true,bookmarksopenlevel=1,
breaklinks=false,pdfborder={0 0 0},backref=false,colorlinks=true]{hyperref}
\hypersetup{citecolor=blue,filecolor=blue,linkcolor=blue,urlcolor=blue}
\usepackage[usenames,dvipsnames]{xcolor}
\usepackage{amssymb,amsmath,mathtools,mathrsfs,enumitem}
\usepackage{epsfig,subfigure,placeins,float}
\usepackage{booktabs,longtable,multirow}
\usepackage{exscale,relsize}
\usepackage[normalem]{ulem}
\usepackage[T1]{fontenc}
\usepackage[utf8]{inputenc}
\usepackage{enumerate}
\usepackage{times, mathptmx}
\usepackage{tikz}
\usepackage{aas_macros}
\usetikzlibrary{arrows,positioning,decorations.markings,decorations.pathmorphing,calc}

\newcommand{\be}{\begin{equation}}
\newcommand{\ee}{\end{equation}}
\newcommand{\pa}{\partial}

\newcommand{\MPl}{M_{\rm Pl}}
\newcommand{\Oo}{{{\cal O}(1)}}

\newcommand{\aB}{\alpha_B}
\newcommand{\aM}{\alpha_M}
\newcommand{\aT}{\alpha_T}
\newcommand{\aK}{\alpha_K}
\newcommand{\aI}{\alpha_i}

\newcommand{\comment}[1]{}

\newcolumntype{C}[1]{>{\centering\let\newline\\\arraybackslash\hspace{0pt}}m{#1}}

\definecolor{hyperref}{RGB}{026,028,087}
\def\gsim{ \lower .75ex \hbox{$\sim$} \llap{\raise .27ex \hbox{$>$}} }
\def\lsim{ \lower .75ex \hbox{$\sim$} \llap{\raise .27ex \hbox{$<$}} }

\def\nn{\nonumber}

\graphicspath{{./figures/}}
\allowdisplaybreaks

\tikzstyle{vecArrow} = [thick, decoration={markings,mark=at position
1 with {\arrow[semithick]{open triangle 60}}},
double distance=1.4pt, shorten >= 5.5pt,
preaction = {decorate},
postaction = {draw,line width=1.4pt, white,shorten >= 4.5pt}]

\begin{document}
\title{Constraining dark energy with the integrated Sachs-Wolfe effect}

\author{Emeric Seraille}
\affiliation{Institute of Cosmology \& Gravitation, University of Portsmouth, Portsmouth, PO1 3FX, U.K.}
\affiliation{\'Ecole Normale Sup\'erieure, 45 Rue d’Ulm, F-75230 Paris, France}
\author{Johannes Noller}
\affiliation{Department of Physics \& Astronomy, University College London, London, WC1E 6BT, U.K.}
\affiliation{Institute of Cosmology \& Gravitation, University of Portsmouth, Portsmouth, PO1 3FX, U.K.}
\affiliation{DAMTP, University of Cambridge, Wilberforce Road, Cambridge, CB3 0WA, U.K.}
\author{Blake D. Sherwin}
\affiliation{DAMTP, University of Cambridge, Wilberforce Road, Cambridge, CB3 0WA, U.K.}

\begin{abstract}
We use the integrated Sachs-Wolfe (ISW) effect, by now detectable at $\sim 5\sigma$ within the context of $\Lambda{}$CDM cosmologies, to place strong constraints on dynamical dark energy theories. 
Working within an effective field theory framework for dark energy we find that including ISW constraints from galaxy-CMB cross-correlations significantly strengthens existing large-scale structure constraints, yielding bounds consistent with $\Lambda{}$CDM and approximately reducing the viable parameter space by $\sim 70\%$. This is a direct consequence of $\Oo$ changes induced in these cross-correlations by otherwise viable dark energy models, which we discuss in detail.
We compute constraints by adapting the $\Lambda{}$CDM ISW likelihood from \cite{Stolzner:2017ged} for dynamical dark energy models using galaxy data from 2MASS, WISE $\times$ SuperCOSMOS, SDSS-DR12, QSOs and NVSS, CMB data from Planck 18, and BAO and RSD large scale structure measurements from BOSS and 6dF.
We show constraints both in terms of EFT-inspired $\alpha_i$ and phenomenological $\mu/\Sigma$ parametrisations. %
Furthermore we discuss the approximations involved and related aspects of bias modelling in detail and highlight what these constraints imply for the underlying dark energy theories.  
\end{abstract}

\date{\today}
\maketitle

\tableofcontents

\section{Introduction} \label{sec-intro}
While dark energy contributes approximately 70\% of the Universe's effective energy budget \cite{Planck:2018vyg}, its precise nature still eludes us. However, there has recently been significant progress in continuously zeroing in more and more on constraining and therefore understanding dark energy. Doing so inevitably takes many forms, ranging from the investigation of specific `lamppost' models to more general parametrised approaches that aspire to be model-independent. One particularly promising avenue in this context have been effective field theory approaches to parametrising dark energy (henceforth EFTDE) that e.g. construct general dark energy interactions assuming dark energy can be modelled as a scalar field and using fundamental theoretical principles to restrict the nature of these interactions \cite{Gubitosi:2012hu,Bloomfield:2012ff,Gleyzes:2013ooa,Gleyzes:2014rba,Bellini:2014fua,Lagos:2016wyv,Lagos:2017hdr}.\footnote{Restricting the new dark energy physics to a single scalar field is motivated by the fact that, if dark energy is indeed dynamical in nature, it generically comes in the form of new particles/fields. So this is a minimal choice in that sense.} 
At the linear level this results in a small number of parameters (really: time-dependent functions) controlling the dynamics of dark energy perturbations imprinted on the large scale structure of the Universe. Notably, when working to second order in derivatives, this neatly matches the linearised version of Horndeski scalar-tensor theories \cite{Bellini:2014fua,Kobayashi:2011nu}, so working with linearised Horndeski or a (linearised) EFTDE is equivalent in this specific context. The small number of parameters controlling linear physics in these theories then is sufficiently rich to capture a wide range of interesting phenomenology, yet suitably restricted so that meaningful constraints on these parameters can be derived from observations. With these EFT/Horndeski-inspired parametrisations having a firm theoretical footing by construction, placing data-driven constraints on them then indeed allows us to zero in on a theoretically well-motivated parameter space of dark energy theories. 

Given its cosmological nature, large scale structure data sets are the most obvious place to turn to for deriving observational constraints on dark energy. Within the context of EFTDE approaches, some of the tightest observational constraints have been obtained e.g. using CMB data, baryon acoustic oscillations, weak-lensing, and redshift space distortions -- see \cite{Noller:2018wyv,Bellini:2015xja,Hu:2013twa,Raveri:2014cka,Gleyzes:2015rua,Kreisch:2017uet,Zumalacarregui:2016pph,Alonso:2016suf,Arai:2017hxj,Frusciante:2018jzw,Reischke:2018ooh,Mancini:2018qtb,Brando:2019xbv,Arjona:2019rfn,Raveri:2019mxg,Perenon:2019dpc,Frusciante:2019xia,SpurioMancini:2019rxy,Bonilla:2019mbm,Baker:2020apq,Joudaki:2020shz,Noller:2020lav,Noller:2020afd,Gsponer:2021obj} for related constraints. 
For specific scalar-tensor theories within the Horndeski class it has also been shown that galaxy-CMB temperature cross-correlations, i.e. the integrated Sachs-Wolfe effect (ISW), can significantly improve observational constraints on such models, e.g. for Kinetic Gravity Braiding models \cite{Kimura:2011td}, for the covariant Galileon \cite{Renk:2017rzu}, and for Galileon Ghost Condensates, Generalized Cubic Covariant Galileons and K-mouflage models \cite{Kable:2021yws}. 
Here we therefore investigate how incorporating galaxy-CMB temperature cross-correlations into the analysis affects constraints at the level of general EFTDE/Horndeski parametrisations. We do so by adapting the $\Lambda{}$CDM ISW likelihood from \cite{Stolzner:2017ged} for dynamical dark energy models (incorporating measurements of galaxy-CMB cross-correlations from 2MASS, WISE $\times$ SuperCOSMOS, SDSS-DR12, QSOs and NVSS)) and combining it with the large scale structure likelihoods used in \cite{Noller:2018wyv} (temperature and polarisation data from Planck 18, baryon acoustic oscillations from BOSS, and redshift space distortions from 6dF and BOSS). Notably (and differently to some of the aforementioned ISW constraints in the dynamical dark energy context), we perform a full Markov chain Monte Carlo (MCMC) exploration of the parameter space and derive the resulting constraints. 

This paper is structured as follows: In section \ref{sec-Horn} we briefly review the relevant aspects of Horndeski gravity and EFTDE approaches. In section \ref{sec-ISW} we then recall the way in which the ISW effect manifests itself in galaxy-CMB cross-correlations, how bias can be modelled and calibrated and what approximations can be used in evaluating the resulting expressions. We pay particular attention to what changes in the evaluation when going from a $\Lambda{}$CDM cosmology to the dynamical dark energy scenarios encoded by EFTDE approaches.  We give details on the data sets used and the likelihood implementation in section \ref{sec-data}. The main results, i.e. cosmological parameter constraints, are presented in section \ref{sec-cosmo}. Here we also present a number of cross-checks and qualitative constraints that provide helpful intuition for how precisely the ISW constraints arise. Finally, we conclude in section \ref{sec-conclusions} and collect additional details in the appendices. 
\\

\section{EFTs of dark energy and Horndeski gravity} \label{sec-Horn}

\subsection{Horndeski gravity}
As discussed in the introduction, modelling dark energy as a scalar field is a minimal choice when considering dynamical dark energy models, in the sense that such models generically introduce novel degrees of freedom/fields. Assuming that a single such degree of freedom (encoded in the scalar) dominates the evolution then is a minimal choice in this context. Horndeski gravity \cite{Horndeski:1974wa,Deffayet:2011gz},\footnote{For the equivalence between the formulations of \cite{Horndeski:1974wa} and \cite{Deffayet:2011gz}, see \cite{Kobayashi:2011nu}.} is a very general framework when investigating the resulting scalar-tensor theories, describing the most general Lorentz-invariant ST action that gives rise to second order equations of motion. 
Here we will for simplicity focus on the subset of such scalar-tensor theories where gravitational waves propagate with the speed of light.\footnote{This trivially satisfies existing observational constraints \cite{TheLIGOScientific:2017qsa,2041-8205-848-2-L14,2041-8205-848-2-L15,2041-8205-848-2-L13,2041-8205-848-2-L12, Jimenez:2015bwa} -- c.f. \cite{Creminelli:2017sry,Sakstein:2017xjx,Ezquiaga:2017ekz,Baker:2017hug} and references therein -- but we also point the interested reader to \cite{deRham:2018red,LISACosmologyWorkingGroup:2022wjo,Baker:2022eiz,Harry:2022zey} 
for more detailed discussions of the (subtle) connection between this assumption and experimental constraints, as well as of models beyond \eqref{Horndeski_simple} consistent with existing constraints.} 
The resulting Lagrangian ${\cal L}_i$ for a scalar $\phi$ and a massless tensor $g_{\mu\nu}$ is described by
\begin{align} \label{Horndeski_simple}
{\cal L}=\Lambda_2^4 G_2(\phi,X) + \frac{\Lambda_2^4}{\Lambda_3^3}G_3(\phi,X)\Box\phi + \MPl^2 G_4(\phi) R,
\end{align}
where $X = -\tfrac{1}{2}\nabla_\mu \phi \nabla^\mu \phi/\Lambda_2^4$ is the scalar kinetic term, and the $G_i$ are free dimensionless functions of $\phi/\MPl$ and $X$. 
Note that we effectively have three mass scales: $\MPl, \Lambda_2$ and $\Lambda_3$. In cosmology they are conventionally taken to satisfy $\Lambda_2 = \MPl H_0$ and $\Lambda_3 = \MPl H_0^2$, which ensures that all interactions can give $\Oo$ contributions to the background evolution.
\\

\subsection{Linear cosmology}
In this paper we will focus on investigating and constraining dark energy phenomenology on the largest scales, well described by cosmological perturbation theory. Treating \eqref{Horndeski_simple} as a candidate description of dark energy on those scales, we are therefore particularly interested in its linearised perturbations around a cosmological FLRW background. Here we will assume this to be a $\Lambda{}$CDM background (motivated by the observed proximity to such a solution) and focus on constraining the behaviour of perturbations around this. 
The freedom in the dynamics of such perturbations is primarily controlled by two functions of time, $\aM$ and $\aB$. More specifically, they are given by \cite{Bellini:2014fua}
\begin{align}
\frac{M^2}{\MPl^2} \aM &= 2 \frac{\dot\phi}{H} G_{4,\phi} , &\frac{M^2}{\MPl^2}\aB &= - 2\frac{\dot{\phi}}{H}\left(XG_{3,X}+ G_{4,\phi}\right).
\label{alphadef}
\end{align}
Here $G_{i,\phi}$ and $G_{i,X}$ denote the partial derivatives of the $G_i$ (with respect to these dimensionless arguments) and $M^2 \equiv 2 \MPl^2 G_4$ is the effective Planck mass seen by tensor perturbations.\footnote{Note that there is a (conventional) sign difference for $G_3$ in the expressions for the $\alpha_i$ here compared to \cite{Bellini:2014fua}. Differences in factors of $H_0$ and/or $\MPl$ are due to our dimensionless definition of the $G_i$ (as opposed to the dimensionful $G_i$ in \cite{Bellini:2014fua}).} $\aM$ is also known as the ``running'' of the effective Planck mass $M$ and $\aB$ as the ``braiding'' that quantifies kinetic mixing between the metric and scalar perturbations in Jordan frame (the frame the $\aI$ are implicitly defined in). Note that there are three other relevant functions of time: 1) the Hubble scale $H$, which encapsulates the background evolution, 2) $\aT$, the difference between the speed of gravitational waves and the speed of light, which is zero here by definition (i.e. by virtue of working with \eqref{Horndeski_simple}), and 3) the `kineticity' $\aK$, which does not affect constraints on other parameters at leading order at the level of linear perturbations  \cite{Bellini:2015xja,Alonso:2016suf}. 

Note that, while here we have summarised the route from an underlying covariant theory to the parameters governing linear perturbations around a cosmological background for this theory, the resulting perturbation theory need not arise out of Horndeski theory and can be seen as an effective description of linear perturbations on a cosmological background in its own right. As discussed above, the same such description arises in different effective field theory approaches \cite{Gubitosi:2012hu,Bloomfield:2012ff,Gleyzes:2013ooa,Gleyzes:2014rba,Bellini:2014fua,Lagos:2016wyv,Lagos:2017hdr}, which construct general interactions of a (perturbative) scalar degree of freedom (in practice: the Goldstone boson associated with the breaking of time translation invariance in cosmology) on an FLRW background.\footnote{This applies when working up to second order in derivatives. When working to higher orders in (derivatives) of perturbations, the results naturally recover perturbative regimes arising out of higher derivative parent covariant theories.} While we will work in the $\aI$ basis outlined above here, our results can therefore straightforwardly be mapped onto different EFTDE formulations -- for notational differences between the two approaches we refer to the summary tables detailing the mapping between different notational choices in \cite{Bellini:2014fua}.

Fully specifying the dynamics of linear cosmological perturbations in the context outlined above requires us to be more explicit about the precise parametric form of the relevant free functions. For a full fiducial underlying Horndeski theory this would require specifying the functional form of the $G_i$, but in the EFT spirit we are adopting here we will instead directly parameterize the $\alpha_i$ that control the dynamics of linear perturbations. While numerous parametrisations exist, here we will focus on arguably the most commonly used parametrisation 
\begin{align}
    \alpha_i = c_i \Omega_{\rm DE},
    \label{alpha_param}
\end{align}
where all $\alpha_i$ are proportional to the fractional dark energy density $\Omega_{\rm DE}$ which determines their time dependence, while all the additional functional freedom is condensed into a single proportionality constant $c_i$ for each given $\alpha_i$. We refer to \cite{Bellini:2014fua,Bellini:2015xja,Linder:2015rcz,Linder:2016wqw,Denissenya:2018mqs,Lombriser:2018olq,Gleyzes:2017kpi,Alonso:2016suf,Noller:2018wyv,Noller:2020afd,Traykova:2021hbr} for a more detailed discussion of the relative merits of different such parametrisations.

\subsection{The Quasi-Static Approximation} \label{subsec-QSA}

The vast majority of scales that we have observational access to are sub-horizon. 
As a result we can employ the quasi-static approximation (QSA) to accurately recover the full evolution of gravitational perturbations \cite{Sawicki:2015zya}.\footnote{
Note that the accuracy of the QSA can be related to the proximity of the background evolution to that of $\Lambda{}$CDM \cite{delaCruz-Dombriz:2008ium,Noller:2013wca}.}
The QSA then amounts to assuming $|\dot{X}| \sim {H} |X|\ll |\partial_i X|$ for any gravitational perturbation field $X$, and hence to a sub-horizon regime where time derivatives of gravitational perturbations can be neglected in comparison to their spatial derivatives. 

Using the QSA we can significantly simplify the Einstein equations at the level of linear perturbations. We work in Newtonian gauge and focus on scalar perturbations, so we have
\begin{equation}\label{Def4Pert}
ds^2=-\left(1+2\Phi\right)dt^2+ a^2\left(1-2\Psi\right)dx_i dx^i,
\end{equation}
where $\Phi$ and $\Psi$ are the standard Bardeen potentials. Manipulating the perturbed Einstein equations derived from \eqref{Horndeski_simple} and applying the QSA, one can now obtain a simple effective Poisson equation as well as an analogous equation for the evolution of the `lensing potential' $\Phi_L \equiv \Phi + \Psi$
\begin{align}
\frac{k^2}{a^2}\Phi &=-4\pi G\mu(a) \rho\Delta, 
&\frac{k^2}{a^2}\Phi_L &=-8\pi G \,\Sigma(a)\rho \Delta,
\label{mrs}
\end{align}
where $\Delta$ is the comoving gauge invariant density perturbation, $G$ is Newton's constant satisfying $G=1/(8\pi M_P^2)$, and we have ignored any anisotropic stress source in the matter sector.
Here $\mu$ quantifies modifications to the effective strength of gravity (i.e. an effective different Newton's constant) and $\Sigma$ parametrises modifications to the effective lensing potential (also sometimes called the Weyl potential) probed by gravitational lensing. GR corresponds to the case when $\mu,\Sigma$ are both unity, so deviations away from unity for either of these parameters are probing the presence of new gravitational physics.\footnote{Sometimes a third parameter is defined in this context: $\gamma$, which measures the presence of an effective gravitational anisotropic stress (often called gravitational ``slip''). This is simply related to the above parameters by $\Sigma=\frac{1}{2}(1+\gamma)\mu$.
}

The $\mu$ and $\Sigma$ functions can explicitly be linked to the $\alpha_i$ that control the evolution of linear perturbations \eqref{alphadef}. Doing so, at leading order one finds\footnote{Note that, even if we had kept $\aK$ in our analysis, it drops out of these expressions at leading order \cite{Bellini:2015xja,Alonso:2016suf}, justifying its omission a posteriori.}
\begin{align}
M^2\mu &= \frac{2\beta_3}{2\beta_1+\beta_2(2-\aB)},
&M^2 \Sigma &=\frac{\beta_1+\beta_2+\beta_3}{2\beta_1+\beta_2\left(2-\aB\right)}. 
 \; 
\label{muSigma}
\end{align}
where we have followed \cite{Alonso:2016suf,Lagos:2017hdr} and introduced the $\beta_{i}$ shorthand functions
\begin{align}
\beta_1 &\equiv - \frac{3(\rho_{\rm tot} + p_{\rm tot})}{H^2M^2} - 2\frac{\dot{H}}{H^2} + \frac{\tfrac{d}{dt}{(\aB H)}}{H^2}, \nn \\
\beta_2 &\equiv \aB + 2\aM, \quad\quad\quad \beta_3 \equiv \beta_1 + (1+\aM)\beta_2,
\label{betas}
\end{align} 
where $\rho_{\rm tot}$ and $p_{\rm tot}$ are the total energy density and pressure in the Universe, and we note we have set $8\pi G = 1$ (and re-scaled all densities and pressures by a factor of 3, using CLASS conventions \cite{Blas:2011rf}).
In the following section we will find that working with the intermediate $\mu,\Sigma$ functions greatly simplifies the evaluation of galaxy-CMB cross-correlations, so the quasi-static expressions summarised here will be particularly useful for us in this context.

\section{Integrated Sachs-Wolfe effect} \label{sec-ISW}

\begin{figure}
    \centering
    \includegraphics[width=0.49\textwidth]{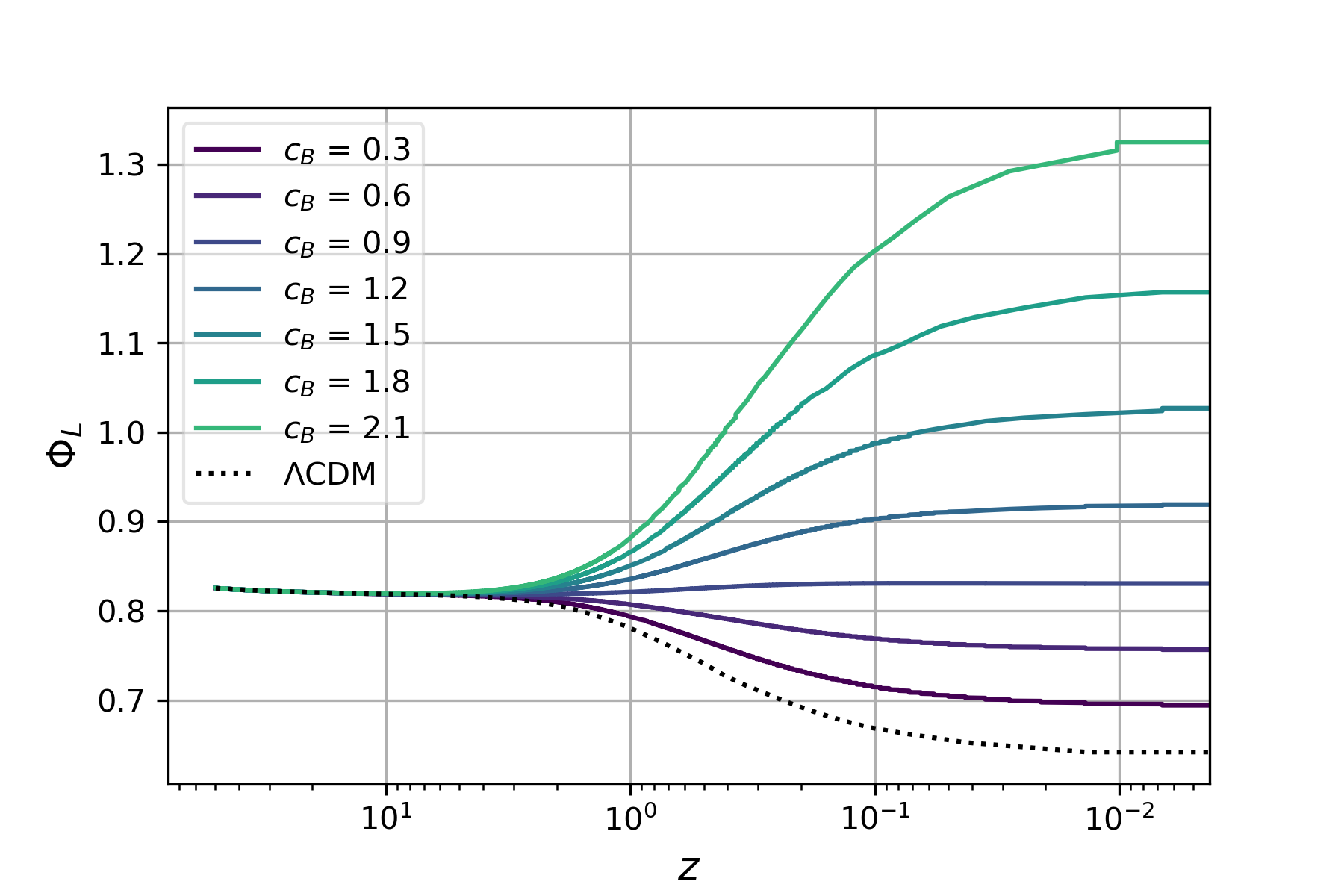}
    \caption{Evolution of the lensing potential $\Phi_L=\Phi+\Psi$ as a function of redshift for various choices of the dark energy parameter $c_b$, where $\alpha_i=c_i \Omega_{DE}$ -- see equation \eqref{alphadef}. All $\Lambda{}$CDM parameters are fixed to the P18 best-fit cosmology \cite{Planck:2018vyg} and in this plot we show results for a fiducial choice of the second dark energy parameter $c_m = 0$. We can clearly see the standard $\Lambda{}$CDM prediction of a decaying lensing potential $\Phi_L$, but also that $\Phi_L$ can both decay or grow in the presence of non-trivial dark energy dynamics. Note that, for such cosmologies, $\Phi_L$ can in principle also decay faster than in $\Lambda{}$CDM, but this requires $c_m \neq 0$ so cannot be seen in this plot, but see figure \ref{fig-Clevolution-v2} for comparison. 
    }
    \label{fig-lensingPotential}
\end{figure}

\begin{figure}
    \centering
    \includegraphics[width=0.49\textwidth]{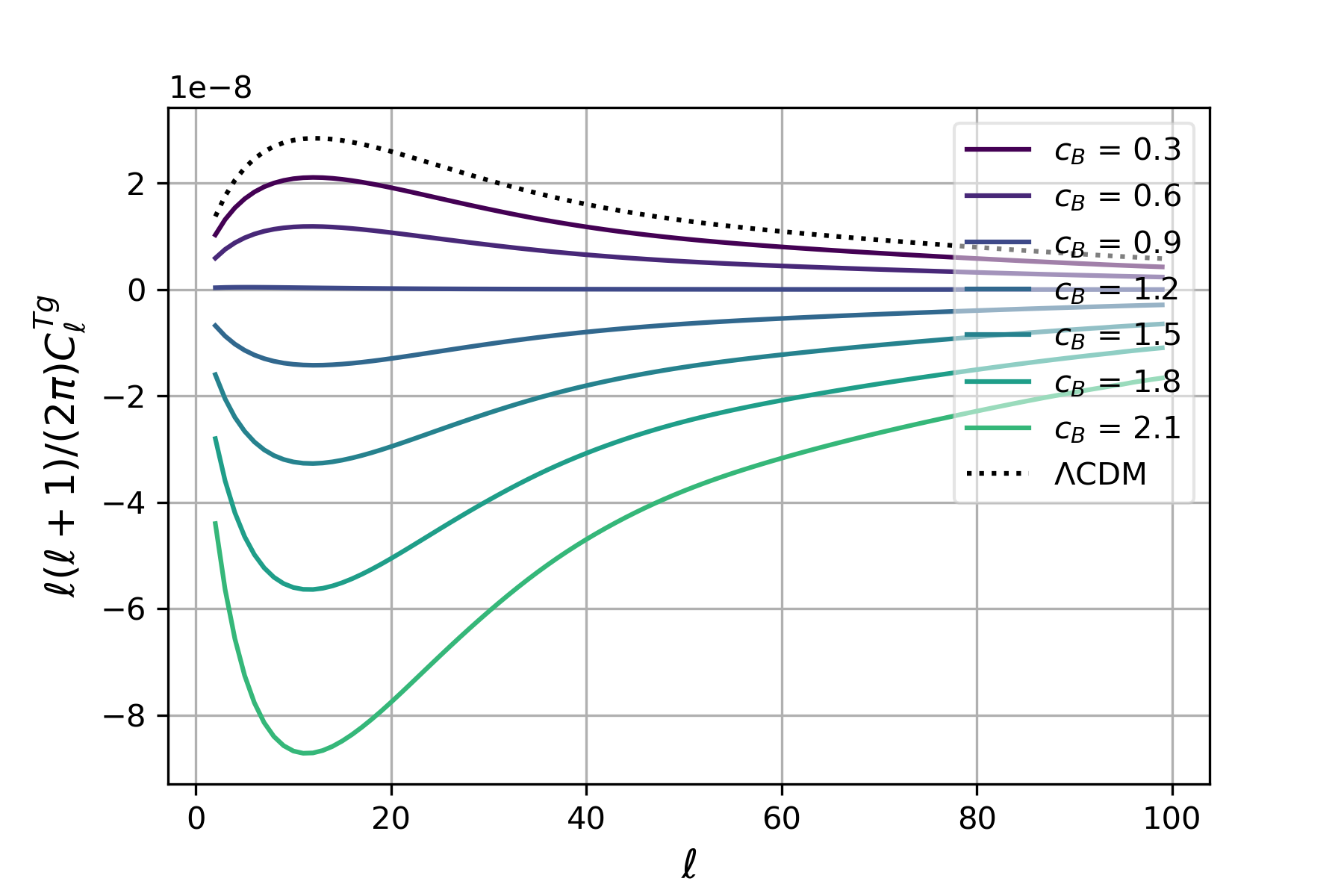}
    \caption{Here we show the $C_\ell^{Tg}$ quantifying cross correlations between galaxies and CMB temperature measurements. As in figure \ref{fig-lensingPotential} above, we vary $c_b$ while keeping $c_m = 0$ fixed. We can neatly see that a decaying lensing potential leads to positive cross-correlations between galaxies and CMB temperature and vice versa, as expected and discussed more explicitly in section \ref{sec-ISW}. The strong $\Oo$ change in the $C_\ell^{Tg}$ signal produced by different models (as shown here) ultimately gives rise to the powerful ISW constraints we find below. the Note that, in order to compute the cross-correlations shown in this figure, we need to choose a fiducial galaxy density distribution -- here we have chosen the distribution from the SDSS-DR12 photo-z sample \cite{Beck_2016}, but the qualitative behaviour of $C_\ell^{Tg}$ is robust under changes of this choice.  
    }
    \label{fig-Clevolution}
\end{figure}

The Integrated Sachs-Wolfe (ISW) effect arises due to the time variation of the gravitational potentials along the lines of sight of CMB photons. This gives rise to (secondary) CMB temperature anisotropies via
\begin{align}
\frac{\delta T_{ISW}(\hat{n})}{\langle T\rangle} = \int_{0}^{z^{*}} dz \; e^{-\tau(z)} \frac{\pa{\Phi_L}}{\pa{z}}(z,\hat{n}\chi(z))
\label{ISWterm}
\end{align}
Where $\hat{n}$ is a unit vector along the line of sight of an observer, $\tau$ is the optical depth, $z$ is redshift, $z^{*}$ is its value at recombination, and $\chi(z)$ is the comoving distance to redshift $z$. During matter-domination $\pa\Phi_L/\pa z  = 0$, so this effect is a natural probe of the dark energy-induced evolution of $\Phi_L$ at late times and hence of dark energy itself. The underlying physics is straightforward: On their way towards an observer, CMB photons will, e.g., fall into the potential well associated with a galaxy and gain energy. Having traversed the potential well the photons will climb back out of the well. If $\Phi_L$ has not evolved in the meantime (as during matter domination), there will be no net gain or loss of energy. However, if $\Phi_L$ has decayed as it does in $\Lambda{}$CDM, then the photon will gain energy leading to a positive correlation of CMB temperature anisotropies with the positions of galaxies. Naturally the opposite, i.e. an anti-correlation, will be the consequence of a growing $\Phi_L$. 

Since dark energy only becomes a main driver of cosmic evolution for $z \lesssim 1.5$, the ISW effect primarily affects large scales and so e.g. affects the CMB temperature power spectrum for large angular scales. While this effect is only ${\cal O}(10\%)$ for $\Lambda{}$CDM and is hence swamped by cosmic variance in that case \cite{Huffenberger:2004tv}, it nevertheless already places powerful constraints on models of dynamical dark energy or modified gravity, where it can be enhanced significantly -- see figure 3 in \cite{Noller:2018wyv} for an explicit illustration of this point. However, as suggested above, a more powerful probe of the ISW effect comes from cross-correlations of CMB temperature with galaxies or quasars (or indeed any other tracer of the underlying gravitational potentials) - see a variety of past ISW analyses employing various tracers of the underlying matter distribution \cite{Afshordi:2003xu,Boughn:2003yz,Fosalba:2003ge,Vielva:2004zg,Cabre:2007rv,Cabre:2006qm,Giannantonio:2006du,Pietrobon:2006gh,McEwen:2006my,Raccanelli:2008bc,Rassat:2006kq,Ho:2008bz,Xia:2009dr,Dupe:2010zs,Hernandez-Monteagudo:2013vwa,Planck:2013owu,Ferraro:2014msa,Planck:2015fcm,Shajib:2016bes,Giannantonio:2008zi,Giannantonio:2012aa,Xavier:2018owe,DES:2018nlb,Calore:2020bpd,Krolewski:2021znk,Kovacs:2021mnf,Hang:2021kfx}. This is what we will exploit here in the context of EFTDE approaches as outlined in the previous section. 
A few points are worth stressing before proceeding:  1) We will assume that $\tau$ is sufficiently small for the relevant redshift range to approximate $e^{-\tau(z)} \sim 1$. This approximation is known to introduce a $\sim 10\%$ error \cite{Stolzner:2017ged}, i.e. below the accuracy with which the ISW effect is typically determined. 2) While $\Phi_L \equiv \Phi + \Psi = 2\Phi$ in $\Lambda{}$CDM, for the dark energy models considered here this need not be the case and indeed generically $\Phi \neq \Psi$. Unsurprisingly, this can strongly affect the ISW signal and, as we will see later on, indeed leads to strong constraints on the $\aI$ dark energy parameters. 3) Note that $\Phi_L$ can grow or decay in an expanding Universe with dynamical dark energy, so the background evolution does not mandate either a growing or decaying $\Phi_L$. 4) When selecting source catalogues for the computation of cross-correlations, focusing on the range $1.5 \gtrsim z \gtrsim 0$ is well motivated given that $\pa\Phi_L/\pa z \sim 0$ during matter domination. In practice, most of the presently available source catalogues come from within this redshift range and so naturally the majority of the data sets we will employ here do as well, but we will use some data points going all the way to $z = 5$ -- see section \ref{sec-data} for details on the precise data sets we will use.

In the same way we present CMB temperature anisotropies due to the ISW effect in equation \eqref{ISWterm}, we can also describe the fractional overdensity of large-scale structure tracers as a function of direction on the sky as follows:
\begin{align}
\frac{\delta N_g(\hat{n})}{\langle N_g\rangle}=\int_{0}^{z^{*}}dz\delta_g(z,\hat{n}\chi(z))\frac{dN_g}{dz}(z) 
\label{galaxyterm}
\end{align}
where $\frac{dN_{g}}{dz}$ represents the normalized selection function for a given kind of compact object tracer (e.g. galaxies, quasars, radio sources) and $\delta_{g}$ is the corresponding density contrast. So while the label $g$ suggests we only apply this to galaxies, this has a wider context in mind. 
Assuming simple linear biasing at each redshift, we can then link this density contrast to the matter density contrast through
\begin{align}
\delta_g(z,\hat{n}\chi(z)) = b_g(z)\delta_{m}(z,\hat{n}\chi(z))
\end{align}
where $\delta_{m}$ is the total matter density contrast and $b_g$ is the bias, which can in general be a function of redshift and depends on the type of compact matter tracer under investigation.
We can now decompose equations \eqref{ISWterm} and \eqref{galaxyterm} on the base of spherical harmonics $Y_{l}^{m}(\theta,\phi)$ as usual, where $\phi$ and $\theta$ are the two angles defining the vector $\hat{n}$, and write
\begin{align}
\frac{\delta T_{ISW}(\hat{n})}{\langle T \rangle} & =\sum_{l,m}a_{l m}^{ISW} Y_{l m}(\hat{n}),
&\frac{\delta N_g(\hat{n})}{\langle N_g \rangle} &= \sum_{l,m}a_{l m}^{g} Y_{l m}(\hat{n}).
\end{align}
Fourier transforming, applying the Rayleigh plane wave identity and isolating the multipole coefficient multiplying each spherical harmonic, we then obtain
\begin{align}
a_{l m}^{ISW} &=-\frac{i^{l}}{2\pi^{2}}\int d^{3}k \int_{0}^{z^{*}} dz e^{-\tau(z)} \frac{\pa{\Phi_L}}{\pa{z}}(z,\vec{k}) j_l[k \chi(z)] Y_{l m}^{*}(\hat{k}), \nn \\
a_{l m}^{g} &= \frac{i^{l}}{2\pi^{2}}\int d^{3}k \int_{0}^{z^{*}} dz b_g(z) \frac{dN_g}{dz}(z) \delta_{m}(k,z) j_l[k\chi(z)] Y_{l m}^{*}(\hat{k}).  
\end{align}
where $k$ is the wavenumber and $j_l$ is the spherical Bessel function of order l. 
Equipped with these expressions, we can write down expressions for the cross-correlation angular power spectrum between CMB temperature and the galaxy field (or field of other compact object tracers) $g$, finding 
\begin{align} \label{eq-ClTg}
C_l^{Tg}=\frac{3 {H_0}^2 \Omega_{m,0}}{\left(D_0(l+\frac{1}{2})\right)^2} \int_{0}^{z^{*}} dz H(z)\frac{dN}{dz}(z)b_g(z){D(z)}^2 A(k_l,z) P(k_l),
\end{align}
where we have defined $A(k,z) \equiv \Sigma(k,z)\bigl(1+(1+z)\frac{d}{dz}ln(D(z)\Sigma(k,z))\bigr)$, the scale $k_l$ satisfies $k_l=(l+\tfrac{1}{2})/\chi(z)$, and $D(z) = (P(k,z)/P(k))^{1/2}$ is the linear growth factor of density fluctuations. 
Note that, in deriving this expression, we have used \eqref{mrs} to solve for the lensing potential and also used the Limber approximation \cite{Limber:1954zz,Kaiser:1991qi,Kaiser:1996tp,LoVerde:2008re,Ho:2008bz}
\begin{align}
\int dk k^{2} h(k) j_l[k\chi] j_l[k \chi^{\prime}] \simeq \frac{\pi}{2}\frac{\delta(\chi-\chi^{\prime})}{\chi^{2}} h\left(\frac{l+\tfrac{1}{2}}{\chi}\right) 
\end{align}
Finally, proceeding along the same lines as above, we can also obtain an expression for the auto-spectrum of the large-scale structure tracer
\begin{align}
   {C_l}^{gg}= \int_{0}^{z^{*}} dz \frac{H(z)}{{(\chi(z)(1+z))}^{2}}\left(\frac{dN}{dz}(z)b(z)D(z)\right)^{2}  P(k_l).
\end{align}

\section{Data sets and likelihood implementation} \label{sec-data}
\begin{table*}[t!] 
\renewcommand{\arraystretch}{1.8}
\setlength{\tabcolsep}{0.3cm}
\begin{tabular}{|l||c|}  \hline  
Catalogue & redshift bins \\ \hline\hline
2MPZ & $[0.01,0.105],[0.105,0.195],[0.195,0.3]$\\ \hline
WISE $\times$ SuperCOSMOS & $[0.09,0.21],[0.21,0.30],[0.30,0.60]$\\ \hline
SDSS-DR12 photo-z & $[0.1,0.3],[0.3,0.4],[0.4,0.5],[0.5,0.7],[0.7,1.0]$
\\ \hline
QSO & $[0.5,1.0],[1.0,2.0],[2.0,3.0]$
\\ \hline
NVSS & $[0,5.0]$\\ \hline
\end{tabular}
\caption{Summary of the source catalogues and corresponding redshift bins from \cite{Stolzner:2017ged}. This is used in our analysis as detailed in section \ref{subsec-galaxy}. 
}
\label{tab_bins}
\end{table*}
Having established the formalism for computing ISW correlations in the previous section, we know briefly summarise the data sets we will use in order to constrain the underlying parameter space. Given that we are focusing on CMB-galaxy cross-correlations, our data sets naturally include several galaxy catalogues as well as CMB measurements and we complement this with additional BAO and RSD measurements.  

\subsection{Source catalogues} \label{subsec-galaxy}
In order to compute galaxy-CMB cross-correlations we adapt the `ISW' likelihood developed for $\Lambda{}$CDM cosmologies by \cite{Stolzner:2017ged}, extending it to be able to handle dark energy and modified gravity theories as outlined in the previous section. The galaxy data sets used are therefore identical to those employed by \cite{Stolzner:2017ged}. They are
the 2MASS photometric redshift catalogue (2MPZ) \cite{Bilicki:2013sza} which samples local large scale structures in the range $z \in [0.00,0.30]$, the WISE $\times$ SuperCOSMOS photo-z catalogue \cite{Bilicki:2016irk}, an all-sky extension of 2MPZ also going to higher redshift $z \in [0.00,0.60]$, the Sloan Digital Sky Survey Data Release 12 (SDSS-DR12) photo-z sample \cite{Beck_2016} with $z \in [0.00,1.00]$ , the catalogue of photometric quasars (QSOs) \cite{Richards:2008eq} tracing  high redshift large scale structures in the range $z \in [0.00,3.00]$, and finally the  NRAO VLA Sky Survey (NVSS) \cite{Condon:1998iy} sampling mostly extragalactic radio sources in the very broad range $z \in [0.00,5.00]$.  %
All the data sets are binned according to redshift, where we summarise all the relevant redshift bins in table \ref{tab_bins}. 
Note that the likelihood in particular uses the ${C_l}^{Tg}$ and ${C_l}^{gg}$ observables discussed in the previous section and the correlation data used are precisely as described in \cite{Stolzner:2017ged}, where they were computed using Planck 2015 CMB maps and masks as well as additional specific masks for each catalogue used -- for further details on the binning as well as on masks applied to these data sets see \cite{Stolzner:2017ged}.
Note that the binning has important implications for the way in which we will model bias, which we discuss in more detail below. 
%

\subsection{Complementary data sets}

We complement the catalogues of galaxies and other compact sources discussed above by CMB temperature, lensing and low-$\ell$ polarisation data from Planck 2018 
\cite{Planck:2015mym,Planck:2015bpv,Planck:2015fie}
Note that working with a $\Lambda{}$CDM background here enables us to work with the {\it Plik\_lite} likelihoods, where most of the internal nuisance parameters are pre-marginalised -- see \cite{Noller:2018wyv} for an explicit justification for this choice.   
As alluded to above, note that it is the late-time ISW effect on the TT power spectrum that is driving CMB constraints on the $\aI$ parameters.
In addition we use baryon acoustic oscillation (BAO) measurements from SDSS/BOSS \cite{Anderson:2014, Ross:2015} as well as redshift space distortion (RSD) measurements from BOSS and 6dF \cite{Beutler:2012, Samushia:2014}. 
Note that there are subtleties related to the observational modelling required in order to extract measurements of $f\sigma_8$ from galaxy surveys, which are particularly relevant in the dynamical dark energy context we are considering here. We will discuss those in more detail below.

\subsection{Bias modelling}\label{subsec-bias}
In the previous section we introduced the redshift-dependent bias $b(z)$, relating overdensities in galaxies (or other sources we are using as matter tracers) to the underlying total matter distribution. In practice the galaxy-galaxy auto-correlations will tightly constrain the bias in a manner that is very robust to small changes in other cosmological parameters. Computing these constraints then necessitates choosing a bias model, with modelling uncertainties arising analogously to our above discussion for the $\aI$ parametrisation. In order to quantify the dependence of our results on those uncertainties in bias modelling and evolution, we will therefore employ two bias models in what follows. 
\begin{itemize}
    \item {\bf Bias model I}: Here we adopt the bias modelling of \cite{Stolzner:2017ged}, which amounts to assigning an independent constant bias factor to each individual bin in all the data sets considered -- see table \ref{tab_bins}. This amounts to 17 bias parameters.
    \item {\bf Bias model II}: Here we adapt the model discussed in \cite{Ferraro:2014msa} and fix the redshift-dependence according to $b_g(z) = b_{\rm survey} (1+z)$ for each survey. Given the five source catalogues we use, this results in 5 bias parameters.
\end{itemize}
As we show in appendix \ref{app-bias}, the second approach produces nearly identical constraints on dark energy and other cosmological parameters, while naturally speeding up the computation of constraints (given the much smaller parameter space dimension).
So while (by design) it does not precisely recover the exact redshift evolution preferred by the data within the context of the more elaborate bias model I, model II nevertheless is an excellent approximation to use here.

\section{Cosmological parameter constraints} \label{sec-cosmo}
 
We are now in a position to place observational constraints on the underlying dark energy parameters. We will do so via a Markov chain Monte Carlo (MCMC) analysis, computing constraints on the modified gravity/dark energy parameters $c_M$ and $c_B$ introduced in \eqref{alpha_param}
while marginalising over the standard $\Lambda{\rm CDM}$ parameters $\Omega_{\rm cdm}, \Omega_{\rm b}, \theta_s,A_s,n_s$ and $\tau_{\rm reio}$.  
Note that we impose a prior $\tau_{\rm reio} \geq 0.04$, corresponding to $z_{\rm reio} \gtrsim 6$, motivated by observations of the Gunn-Peterson trough \cite{SDSS:2001tew}.
Following \cite{Planck:2015fie}, our fiducial model includes two massless and one massive neutrino eigenstate with the minimal sum of masses allowed by oscillation experiments, $\sum_\nu m_\nu=0.06 \; {\rm eV}$.
For related cosmological parameter constraints on deviations from GR using general parameterised approaches and a variety of (current and forecasted) experimental data, see
\cite{Noller:2018wyv,Bellini:2015xja,Hu:2013twa,Raveri:2014cka,Gleyzes:2015rua,Kreisch:2017uet,Zumalacarregui:2016pph,Alonso:2016suf,Arai:2017hxj,Frusciante:2018jzw,Reischke:2018ooh,Mancini:2018qtb,Noller:2018eht,Perenon:2019dpc,Frusciante:2019xia,Arai:2019zul,Scott:2022fev,Andrade:2023pws,Nguyen:2023fip,Wen:2023bcj,Castello:2023zjr,Sugiyama:2023tes}. 
Finally we will work with a fiducial $c_k = 0.1$ choice throughout. As discussed above, the `kineticity' $\alpha_K$ does not affect cosmological constraints at leading order, and indeed it has been explicitly verified that fixing a fiducial $c_k$ as we do here does not bias the constraints on the other $c_i$ \cite{Bellini:2015xja}. For reference throughout the remainder of this section, we adopt the following shorthands in referring to the datasets detailed in the previous section:
\begin{itemize}
\item {P18B}: This prior includes Planck 2018 CMB temperature, CMB lensing and low-$\ell$ polarisation data 
\cite{Planck:2015mym,Planck:2015bpv,Planck:2015fie} as well as baryon acoustic oscillation (BAO) measurements from SDSS/BOSS \cite{Anderson:2014, Ross:2015}
\item {RSD}: Here we include redshift space distortion (RSD) constraints from BOSS and 6dF \cite{Beutler:2012, Samushia:2014}. 
Note that this is a minimal set of RSD constraints chosen to be conservative and eliminate the risk of introducing unwanted correlations between (RSD) data sets -- see \cite{Noller:2018wyv} for details. 
\item {ISW}:  This label denotes the constraints coming from the auto- and CMB cross-correlations computed using the 2MPZ, WISE $\times$ SuperCOSMOS, SDSS-DR12 photo-z, QSO, and NVSS data sets detailed in \ref{subsec-galaxy}. 
\end{itemize} 
\vspace{.2cm}

\subsection{Baseline CMB constraints}
\begin{figure}
    \centering
    \includegraphics[width=0.48\textwidth]{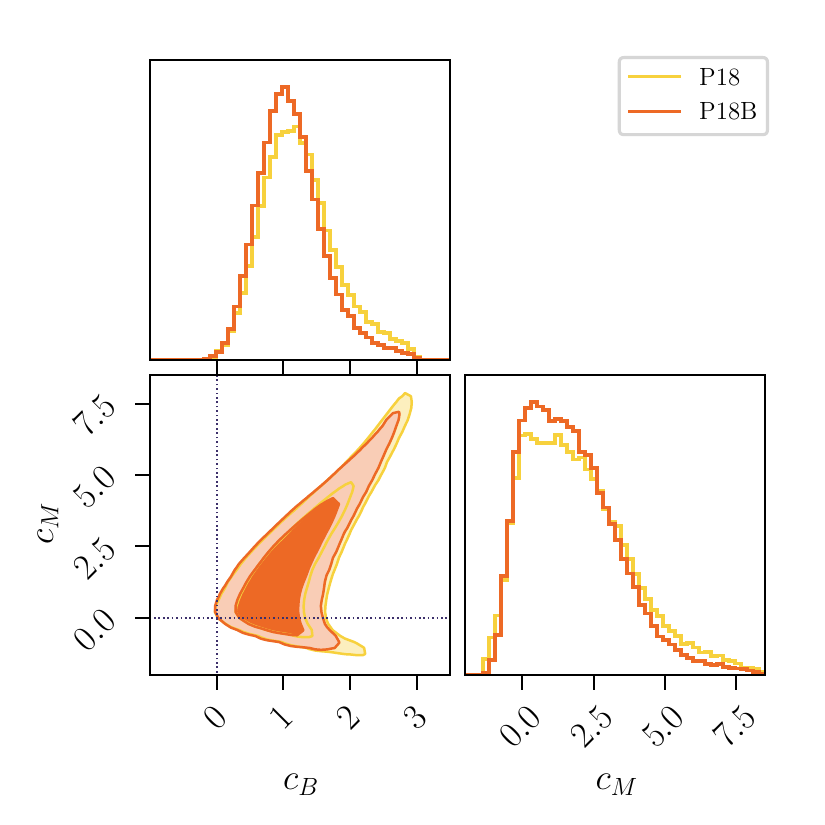}
    \caption{Here we show marginalised constraints for the dark energy $\{c_b,c_m\}$ parameters using Planck 18 (P18) and Planck 18 + BAO data (P18B)  -- for details on the data sets used see section \ref{sec-data}. Contours mark $68\%$ and $95\%$ confidence intervals and the origin corresponds to the $\Lambda{}$CDM limit (here consistent at $\sim 2\sigma$). For the $\Lambda{}$CDM background we are considering here, incorporating BAO data visibly only very mildly affects overall constraints on $\{c_b,c_m\}$ and CMB-only constraints are approximately of the (as we will see: relatively weak) order $c_i \lesssim {\cal O}(10)$. 
    }
    \label{fig-PlanckBaseline}
\end{figure}
As a first step, we briefly recap CMB constraints on the modified gravity/dark energy parameters $c_M$ and $c_B$ -- for details on the derivation of these constraints in the present context of a $\alpha_i = c_i \Omega_{\rm DE}$ parametrisation see e.g. \cite{Bellini:2015xja,Noller:2018wyv}. We summarise these constraints in figure \ref{fig-PlanckBaseline}. As we can see there, CMB-only constraints (more specifically, temperature, lensing and low-$\ell$ polarisation data from Planck 18) already provide significant constraints on the marginalised $\{c_m,c_b\}$ parameter space, to a first approximation enforcing $c_i \lesssim {\cal O}(10)$ without any strong degeneracies in this parameter space. In this figure we also show that the inclusion of BAO constraints only has a minimal effect, consistent with the finding of \cite{Noller:2018wyv} and largely due to the fact that we assume a $\Lambda{}$CDM background and the BAO measurements considered here do not add significant constraining power on top of the CMB data sets detailed above in this context. As such, we will only refer to the combined CMB+BAO constraints (P18B) in the following sections.

\subsection{Qualitative ISW constraints}\label{subsec-qualitativeISW}

\begin{figure}[t!]
\centering
\includegraphics[width=0.48\textwidth]{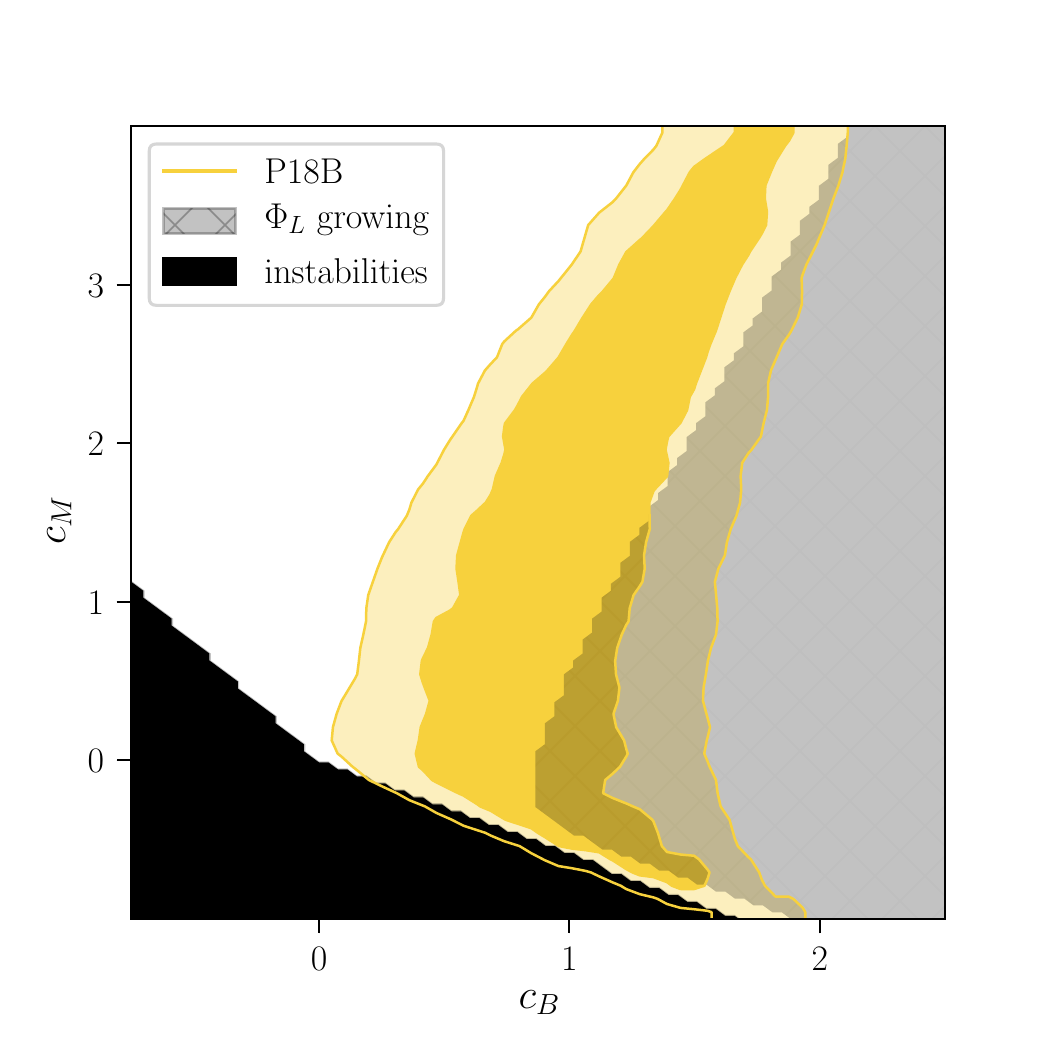}
\caption{Qualitative ISW constraint on the (marginalised) $\{c_b,c_m\}$ plane. We again plot the CMB+BAO constraints from figure \ref{fig-PlanckBaseline}, but here also show how what effectively amounts to a naive ISW prior impacts the parameter space. Black areas are those plagued by gradient instabilities, while grey hashed regions are those where the lensing potential $\Phi_L$ grows (between redshifts $5$ and $0$). Given current constraints within $\Lambda{}$CDM detect the ISW effect, in other words a decaying lensing potential, at $\sim 5\sigma$ \cite{Stolzner:2017ged}, one would expect that the grey hashed regions of parameter space are strongly disfavoured and incorporating full ISW constraints allows placing yet tighter bounds. We will indeed find this to be the case in a more complete analysis below, but the qualitative constraints outlined here already provide useful intuition for the kind of constraints one can expect. 
}
\label{fig-qualitativeISW}
\end{figure}

Before computing the observational constraints from the LSS data sets we detail above (both ISW and RSD), it is worth to consider where and how we expect ISW constraints to act. In other words, to identify which parts of parameter space one should expect to be ruled out/restricted by such constraints prior to any detailed likelihood calculation. To this end it is important to recall that current measurements of the ISW effect establish that there is a positive cross-correlations between CMB temperature anisotropies and (the position of) galaxies at $\sim 5\sigma$ \cite{Stolzner:2017ged}. As discussed in section \ref{sec-ISW}, this is ultimately telling us that the lensing potential has a positive derivative in redshift, i.e. the lensing potential is decaying as a function of time in the late Universe. While this decay is indeed a prediction in standard $\Lambda{}$CDM cosmologies, this is not generically true for the EFTDE theories we are considering here. We show this explicitly in figure \ref{fig-lensingPotential} where we inspect a slice of theory space with $c_m=0$ and varying $c_b$ values broadly consistent with the P18B constraints discussed and shown above in figure \ref{fig-PlanckBaseline}. The $\Oo$ changes to the predicted cross-correlations shown immediately suggest that measurements of the ISW effect will yield powerful constraints on dark energy physics. To calibrate expectations it makes sense to first demarcate regions of parameter space where the lensing potential is indeed decaying vs. regions where it is growing. At the very least we would expect observational constraints to rule out the latter region. We perform this qualitative check in figure \ref{fig-qualitativeISW}.  Comparing with the P18B constraints discussed above 
and reproduced in the figure, it is clear that significant regions of the marginalised $\{c_m,c_b\}$ parameter space that are consistent with CMB-only constraints are inconsistent with a decaying lensing potential in the late Universe and so we would indeed expect incorporating ISW constraints into the analysis to yield powerful constraints on the landscape of dark energy theories investigated here.

\begin{figure*}[ht!]
    \centering
    \includegraphics[width=0.48\textwidth]{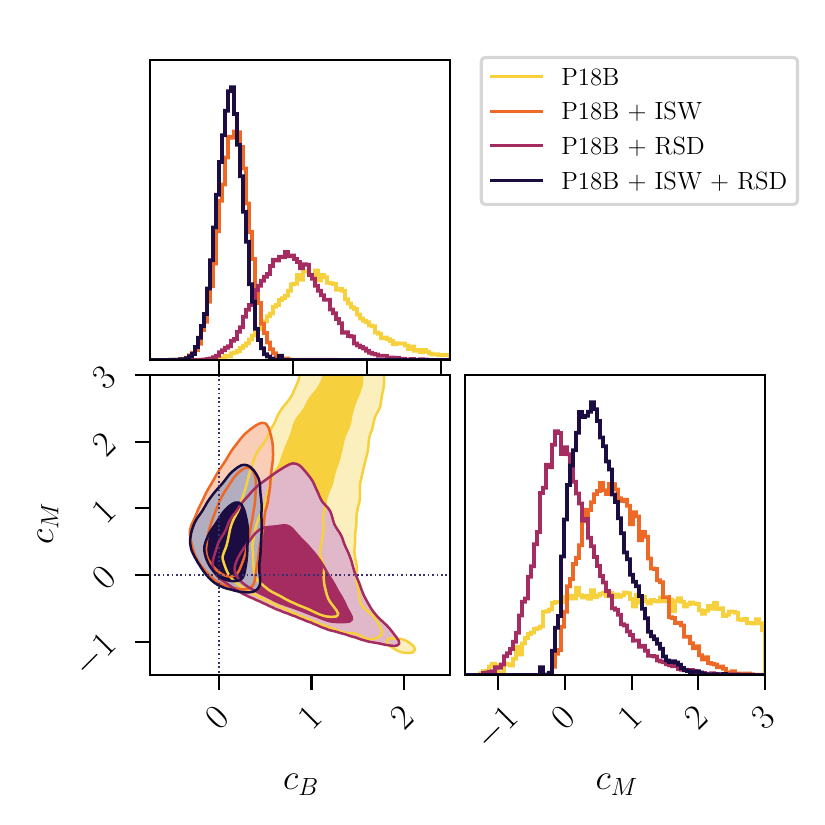}
    \includegraphics[width=0.48\textwidth]{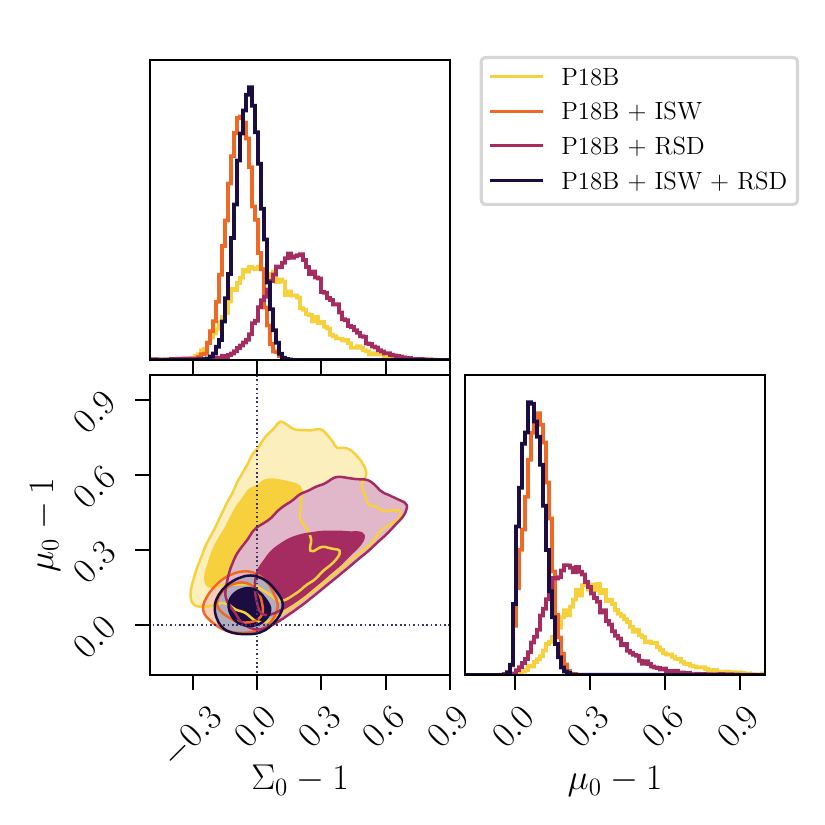}
    \caption{{\bf Left panel}: Here we show marginalised constraints for the dark energy $\{c_b,c_m\}$ parameters for different combinations of data sets. Contours mark $68\%$ and $95\%$ confidence intervals, and are computed using Planck 18 and BAO data (P18B), as well as measurements of galaxy-galaxy auto- and galaxy-CMB temperature cross correlations (ISW), and redshift space distortion measurements (RSD) -- for details on the data sets used see section \ref{sec-data}. Note that ISW constraints effectively enforce $c_b \lesssim 0.5$ and incorporating them drastically improves CMB+BAO-only bounds. RSD measurements in addition impose a somewhat tighter bound on $c_m$, namely $c_m \lesssim 2$, so that a combination of bounds from all data sets used here allows another (more mild improvement) of constraints.
    {\bf Right panel}: Analogous plot to the left panel, 
    but for the more phenomenological $\mu/\Sigma$ parametrisation, where $\mu_0/\Sigma_0$ are derived from the 
    $c_i$/$\alpha_i$ via \eqref{muSigma}. The $0$ subscript denotes the value of $\mu/\Sigma$ today. Projected into the $\mu_0/\Sigma_0$ parameter space, the constraining power of adding ISW measurements is highlighted even more than for the $c_i$ parametrisation in figure \ref{fig-ObservationalConstraints}, with a stark improvement in going from P18B to P18B+ISW constraints, while there is only a marginal difference between P18B+ISW and P18B+ISW+RSD bounds. We emphasise that the strong $\mu_0/\Sigma_0$ constraints found here incorporate a `theoretical prior' of working with an underlying EFT $c_i$ parameter basis -- see main text for a more detailed discussion.
    }
    \label{fig-ObservationalConstraints}
\end{figure*}

\subsection{Observational ISW constraints I: EFT parametrisation}
\label{subsec-mainResults}
Our main results are shown in figure \ref{fig-ObservationalConstraints}. Here we compare constraints in the $\{c_m,c_b\}$ plane obtained via different combination of the data sets detailed above. We see that ISW data in particular rule out $c_b \gtrsim 1$ when combined with P18B constraints. As expected, this is in excellent agreement with the expectation from the qualitative ISW constraints shown in figure \ref{fig-qualitativeISW} and discussed in detail above.
As we stressed there, current measurements establish that there a decaying lensing potential (i.e. a positive cross-correlations between CMB temperature anisotropies and the position of galaxies) at $\sim 5\sigma$ \cite{Stolzner:2017ged}. It is therefore unsurprising that the  ISW constraints shown in figure \ref{fig-ObservationalConstraints} are significantly stronger than the conservative qualitative estimate above -- we show this comparison explicitly in figure \ref{fig-comboPlot}. 
As we discuss in more detail in appendix \ref{app-ISWpriors}, the data constraints rule out additional regions of parameter space where the lensing potential does not decay sufficiently quickly (including cases where it does decay at early and late times, but plateaus at intermediate redshifts).
It is also worth noting that the addition of the ISW-related data sets discussed here, while significantly tightening constraints in the $\{c_m,c_b\}$ plane, only has a very small effect on the other varied (standard $\Lambda{}$CDM) parameters -- see appendix \ref{app-LargeComp} for more details on this point. This is because changing these parameters within existing CMB constraints does not qualitatively alter the evolution of the lensing potential and hence these parameter changes only mildly affect the observables probed by the ISW effect. The ISW measurements used here are therefore a particularly clean and powerful way to constrain the dark energy/modified gravity effects we are focusing on in this paper. 

We also compare and contrast constraints derived from P18B+ISW with those derived using another set of LSS constraints, namely RSD constraints as detailed in \cite{Noller:2018wyv}. Note that these RSD constraints are more subtle in that extracting $f\sigma_8$ measurements from the data sets in question is somewhat cosmology-dependent. 
More specifically, note that observational modeling is required to extract measurements of $f\sigma_8$ from galaxy surveys, which are typically constructed based on GR and validated using GR mock catalogues. Using such standard measurements has been shown to give accurate results e.g. in the context of DGP models \cite{Barreira:2016ovx}, but not in at least some models displaying scale-dependent linear growth \cite{Taruya:2013quf}.
It is therefore worth emphasising that the ISW constraints discussed here are not affected by an analogous modelling proviso and are therefore more robust in this specific sense. 
In terms of the marginalised $\{c_m,c_b\}$ parameter space, RSD constraints in particular constrain $c_m$ to be bounded by $c_m \lesssim 2$. This is intuitive since clustering as measured by RSD constraints is highly sensitive to changes in the effective Newtonian constant $G_{\rm eff}$ and hence to changes in the effective Planck mass as encoded by $c_m$. 
Because of the different and orthogonal ways in which they constrain the parameter space, ISW and RSD constraints are highly complementary and so combining them can significantly improve the overall constraining power. For the minimal RSD constraints shown here, adding RSD to P18B+ISW constraints only mildly improves bounds, but this will change when considering more comprehensive RSD constraints in the future \cite{redshift}.\footnote{Note that going beyond the minimal and conservative RSD data set used here while avoiding using correlated data sets (that can bias constraints) requires careful data selection and modelling.} It is worth pointing out that the inclusion of ISW data also results in constraints that are `more' consistent with $\Lambda{}$CDM, removing the $\sim 2\sigma$ preference for non-$\Lambda{}$CDM parameter values displayed by P18B and P18B + RSD constraints.
Overall we find that the viable region/volume in $c_i$ parameter-space is significantly restricted by including ISW constraints, with that region shrinking by $\sim 70 \% $ when comparing P18B + ISW + RSD constraints with P18B + RSD constraints.\footnote{By the “volume” in $c_i$ parameter space we mean the very simple measure $\Delta c_b \Delta c_m$ , where $\Delta c_i$ denotes the $95\%$ confidence interval for $c_i$ (note that this measure is not unique and many alternative measures exist). For example, for the P18B + ISW + RSD constraints shown in table \ref{tab_ci_bounds} we have $\Delta c_b = 0.28 + 0.29 = 0.57$.}

\begin{table*}[t!] 
\renewcommand{\arraystretch}{1.8}
\setlength{\tabcolsep}{0.3cm}
\begin{tabular}{|l||c|c||c|c|}  \hline  
Data sets & $c_b$ & $c_m$ & $\mu_0 - 1$ & $\Sigma_0 -1$\\ \hline\hline
P18B & $1.17^{+1.27}_{-0.85}$ & $1.87^{+4.09}_{-2.37}$ & $0.38^{+0.45}_{-0.28}$ & $0.07^{+0.42}_{-0.30}$\\ \hline
P18B + ISW & $0.16^{+0.33}_{-0.33}$ & $0.86^{+1.15}_{-0.84}$ & $0.08^{+0.09}_{-0.08}$ & $-0.08^{+0.12}_{-0.13}$\\ \hline
P18B + RSD & $0.84^{+0.74}_{-0.66}$ & $0.13^{+1.26}_{-0.86}$ & $0.27^{+0.30}_{-0.20}$ & $0.21^{+0.36}_{-0.30}$\\ \hline
P18B + ISW + RSD & $0.12^{+0.28}_{-0.29}$ & $0.54^{+0.90}_{-0.60}$ & $0.07^{+0.09}_{-0.07}$ & $-0.04^{+0.11}_{-0.12}$\\ \hline
\end{tabular}
\caption{Here we show $2\sigma$ bounds (marginalised 1D posteriors) on the dark energy/modified gravity $c_i$ parameters for different combinations of data sets, as discussed in section \ref{sec-cosmo} (and especially subsections \ref{subsec-mainResults} and \ref{subsec-muSigmaResults}). We also show the corresponding bounds on $\mu_0$ and $\Sigma_0$, where the $0$ subscript denotes the value of these parameters today and the parameters themselves are inferred from the $c_i$ (and ultimately $\alpha_i$) using the expressions discussed around equation \eqref{muSigma}.}
\label{tab_ci_bounds}
\end{table*}

\subsection{Observational ISW constraints II: \texorpdfstring{$\mu/\Sigma$}{[mu/Sigma]} parametrisation}
\label{subsec-muSigmaResults}

Above we discussed observational constraints in the $\alpha_i$ parameter space when using different data sets. Here we would like to map these constraints into the $\{\mu,\Sigma\}$ parameter space introduced in section \ref{subsec-QSA}. This parameter space is often seen as a more phenomenological alternative to the more theory-informed $\alpha_i$ parametrisation. Indeed, constraints on $\{\mu,\Sigma\}$ are frequently obtained in this way, by parametrising these parameters directly -- see e.g. recent survey results from Planck 2018 and the Dark Energy Survey Y3 in \cite{Planck:2018vyg,DES:2022ccp}. Instead here we always ultimately work with the theory-informed $\alpha_i$ basis, parametrising this directly and hence constraining the corresponding $c_i$ parameter space, but we use the fact that a well-defined mapping exists from this parameter space to $\{\mu,\Sigma\}$ -- as detailed in section \ref{subsec-QSA}. We stress that the choice of implicit (as in our case via the underlying $\alpha_i$ basis) or explicit (as in \cite{Planck:2018vyg,DES:2022ccp}) parametrisation of  $\{\mu,\Sigma\}$  has a significant effect on the precise constraints obtained and one should therefore not expect identical constraints when using different parametrisations. We illustrate this point in appendix \ref{app_proptoa}, where we show analogous constraints to those shown in the main text, but for a $\alpha_i = c_i a$ parametrisation instead of the $\alpha_i = c_i \Omega_{\rm DE}$ parametrisation \eqref{alpha_param} employed in the main text.

In the right panel of figure \ref{fig-ObservationalConstraints} we show the constraints on $\{\mu,\Sigma\}$ implied by our underlying theory-informed $c_i$ constrains as shown in the left panel of the same figure. 
It is instructive to think of this mapping as constituting a theoretical prior impacting the resulting constraints on $\{\mu,\Sigma\}$.\footnote{We therefore re-emphasise that care ought to be taken when comparing these constraints with constraints that are obtained for different, direct parametrisations of $\{\mu,\Sigma\}$ encountered in more phenomenological approaches, since differences in the resulting constraints may be driven by the parametrisations employed rather than e.g. the data sets being used.}  
Figure \ref{fig-ObservationalConstraints} then clearly shows how powerful the addition of ISW constraints is, even more so than when constraints are expressed in the $c_i$ basis. The inclusion of ISW constraints significantly tightens constraints on $\Sigma$, which is intuitive since the (time derivative of the) lensing potential is precisely what the ISW effect probes. c.f. equation \eqref{eq-ClTg}. What is less intuitive is that $\mu$ is likewise strongly constrained, which is (intuitively) more linked to clustering and is therefore indeed significantly constrained when e.g. including RSD measurements, as can also be seen in figure \ref{fig-ObservationalConstraints}. That the inclusion of ISW constraints also places further strong constraints on $\mu$ should therefore be interpreted as a consequence of the `theoretical prior' imposed by working with an underlying $c_i$ parameter basis. Both $\alpha_M$ and $\alpha_B$ affect $\mu$ and $\Sigma$ and are constrained by all the data sets being used here, so strong additional constraints on $\mu$ are a consequence of using this underlying theory-informed parametrisation here. The resulting overall improvement in constraints is remarkable. 
We find that the viable region/volume in $\{\mu_0,\Sigma_0\}$ parameter-space is significantly shrunk by including ISW constraints, namely by close to $\sim 90 \% $ when comparing P18B + ISW + RSD constraints with P18B + RSD constraints.\footnote{In analogy to our previous discussion for the $c_i$ parameter space, by the “volume” in $\mu/\Sigma$ parameter space we mean the very simple measure $\Delta \mu_0 \Delta \Sigma_0$, where $\Delta \mu_0$ denotes the $95\%$ confidence interval for $\mu$ etc.}

\begin{figure}[t!]
    \centering
    \includegraphics[width = 0.48\textwidth]{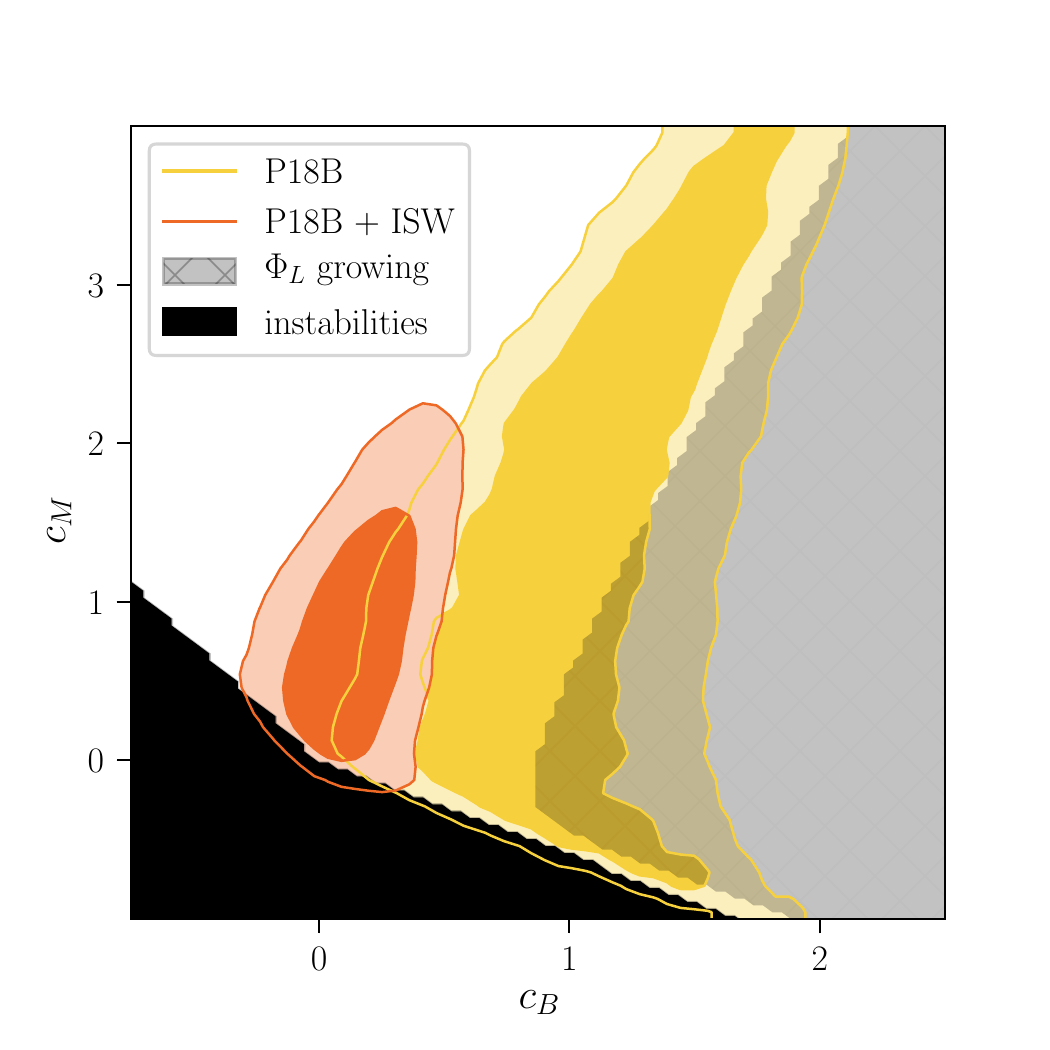}
    \caption{Comparison between the P18B and ISW observational constraints shown in figure \ref{fig-ObservationalConstraints} and the minimal `qualitative' constraints shown in figure \ref{fig-qualitativeISW}.  Black regions are excluded by the presence of (gradient) instabilities, while the hashed out grey area corresponds to a `naive' ISW exclusion prior ruling out regions of parameter space where the lensing potential does not decay (between redshift 5 and 0, although the precise choice of this redshift range has a minimal effect). Since the ISW effect is currently detectable at $\sim 5\sigma$, the data driven ISW constraints are (as expected) significantly stronger than just applying this naive prior to P18B constraints would yield.  
    }
    \label{fig-comboPlot}
\end{figure}

\section{Conclusions} \label{sec-conclusions}
In this paper we have investigated to what extent the ISW effect can be used to tighten the observational constraints we can place on dynamical dark energy theories. Working within an effective field theory formalism that captures the behaviour of scalar perturbations in a wide range of dynamical dark energy/modified gravity theories around a cosmological background space-time, we indeed found that ISW constraints from galaxy-CMB cross-correlations significantly improve constraints. More specifically, our key findings include:
\begin{itemize}
    \item Including ISW constraints reduces the viable parameter space from CMB+BAO+RSD constraints by $\sim 70 \%$ for theoretically motivated `EFT' parametrisations of the underlying dark energy/modified gravity physics. This corresponds to a $\sim 90\%$ improvement when mapped to the frequently-used phenomenological $\mu/\Sigma$ parameters. The constraints we find are consistent with $\Lambda{}$CDM and remove $\sim 2\sigma$ preferences for a departure from $\Lambda{}$CDM found when only employing CMB+BAO+RSD constraints.
    \item  When working with the $\alpha_i$ EFT functions, we find that {ISW measurements in particular constrain the `braiding' function} $\alpha_B$. In terms of the $\alpha_i = c_i \Omega_{\rm DE}$ parametrisation, this quantitatively means $c_b \lesssim 0.5$ at the $2\sigma$ confidence level. 
    \item We show that these constraints {are robust} when employing different bias models (see appendix \ref{app-bias} for details),   
    but {quantitatively depend} on which precise parametrisations are employed for the free functions controlling the dynamics of (linear) perturbations (see appendix \ref{app_proptoa} for details).
\end{itemize}
Going forward, obvious next steps include the incorporation of larger data sets, both in terms of the survey catalogues used to compute ISW cross-correlations as well as for complementary constraints from other large scale structure probes -- in particular see \cite{redshift} for an upcoming discussion of improved constraints from redshift space distortions. 
Finally, note that we have only considered (linear) large scale structure constraints here, implicitly constraining a dark energy EFT for the associated large scales only. When going beyond this, additional subtleties in connecting constraints from a large range of scales arise,  additional constraints can e.g. be found by 1) considering gravitational-wave induced dark energy instabilities \cite{Creminelli:2019kjy}, significantly further tightening cosmological parameter constraints on dark energy \cite{Noller:2020afd}, 2) deriving complementary constraints tightly linking local solar system constraints to cosmology \cite{Babichev:2011iz,Burrage:2020jkj,Noller:2020lav}, 3) demanding that such theories have a sensible UV completion -- see \cite{Melville:2019wyy,Kennedy:2020ehn,deRham:2021fpu,Melville:2022ykg} and references therein for a discussion of how the resulting bounds can affect the cosmological parameter constraints discussed here, 
4) going beyond linear dynamics and using non-linear information to further constrain the dark energy/modified gravity theories considered here, e.g. along the lines discussed in \cite{Cusin:2017mzw,Cusin:2017wjg,Thomas:2020duj,Srinivasan:2021gib,Fiorini:2022srj,Brando:2022gvg,Wright:2022krq,Bose:2022vwi},
5) incorporating additional theoretical priors into the analysis, e.g. related to radiative stability \cite{Noller:2018eht,Heisenberg:2020cyi} or to improved theoretical modelling of the EFT functions \cite{Traykova:2021hbr}. 
All these different probes constrain dark energy-related physics in a variety of complementary and powerful ways and it will be interesting to better understand their interplay in the future. Returning our focus to the large scale structure regime we have investigated in this paper, with a plethora of new large scale structure surveys and the associated data coming up, it will be fascinating to see how constraints on the nature of dark energy evolve in the years to come. What we hope to have shown here is that, in the context of theoretically well-motivated models, CMB/galaxy cross-correlations can play a very significant role in this enterprise.

\section*{Acknowledgments}
\vspace{-0.1in}
\noindent 
JN is supported by an STFC Ernest Rutherford Fellowship (ST/S004572/1). 
In deriving the results of this paper, we have used: CLASS \cite{Blas:2011rf}, corner \cite{corner}, hi\_class \cite{Zumalacarregui:2016pph,Bellini:2019syt}, MontePyton \cite{Audren:2012wb,Brinckmann:2018cvx} and xAct \cite{xAct}.  
\\

\noindent{\bf Data availability} Supporting research data are available on reasonable request from the authors.

\appendix

\section{Cosmological constraints and bias models}
\label{app-bias}

\begin{figure}[t!]
    \centering
    \includegraphics[scale=0.65]{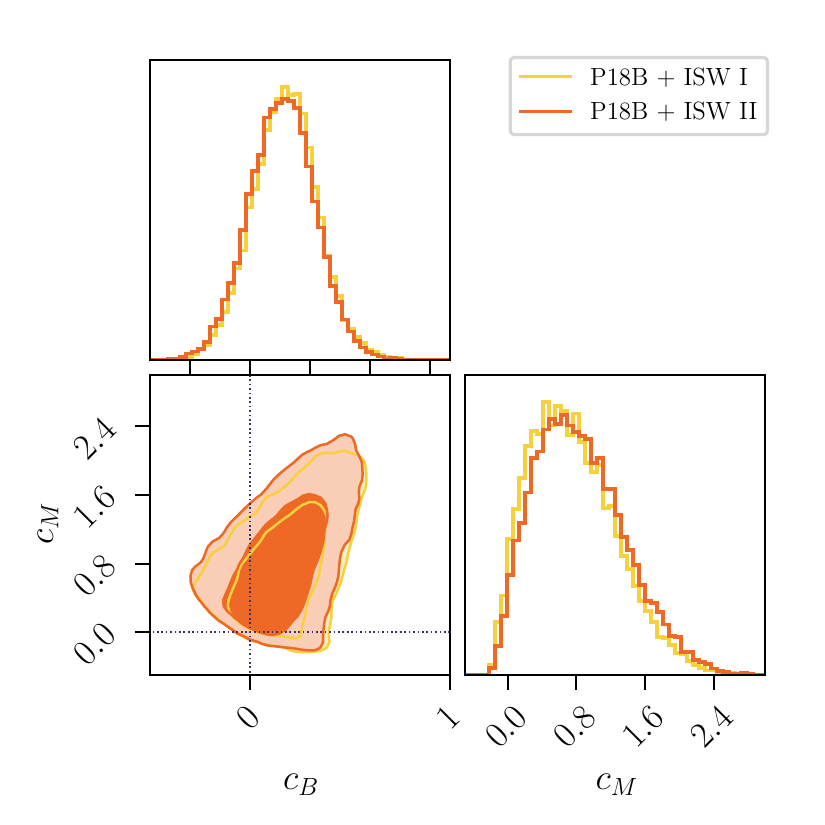}
    \caption{Comparison between constraints for different bias models in the $\{c_m,c_b\}$ plane. Here we use CMB + BAO + ISW data, but employ bias model I and II as detailed in section \ref{subsec-bias}. We can clearly see that the simplified bias model II recovers near-identical constraints on the $\{c_m,c_b\}$ parameters and have checked that this remains true for other cosmological parameters as well.
    }
    \label{fig-biasComp}
\end{figure}

In section \ref{subsec-bias} we introduced two different bias models, one identical to the setup employed by \cite{Stolzner:2017ged} and a second, simplified model with a specific ansatz for the redshift dependence of the bias function and hence fewer bias parameters. In figure \ref{fig-biasComp} we show that the differences between the two bias models do not have a significant effect on cosmological parameter constraints in the $\{c_b,c_m\}$ space (marginalised over all standard $\Lambda{}$CDM parameters). We have also checked that this remains true for the standard $\Lambda{}$CDM parameters themselves.  

\begin{figure}[t!]
    \centering
    \includegraphics[width = 0.48\textwidth]{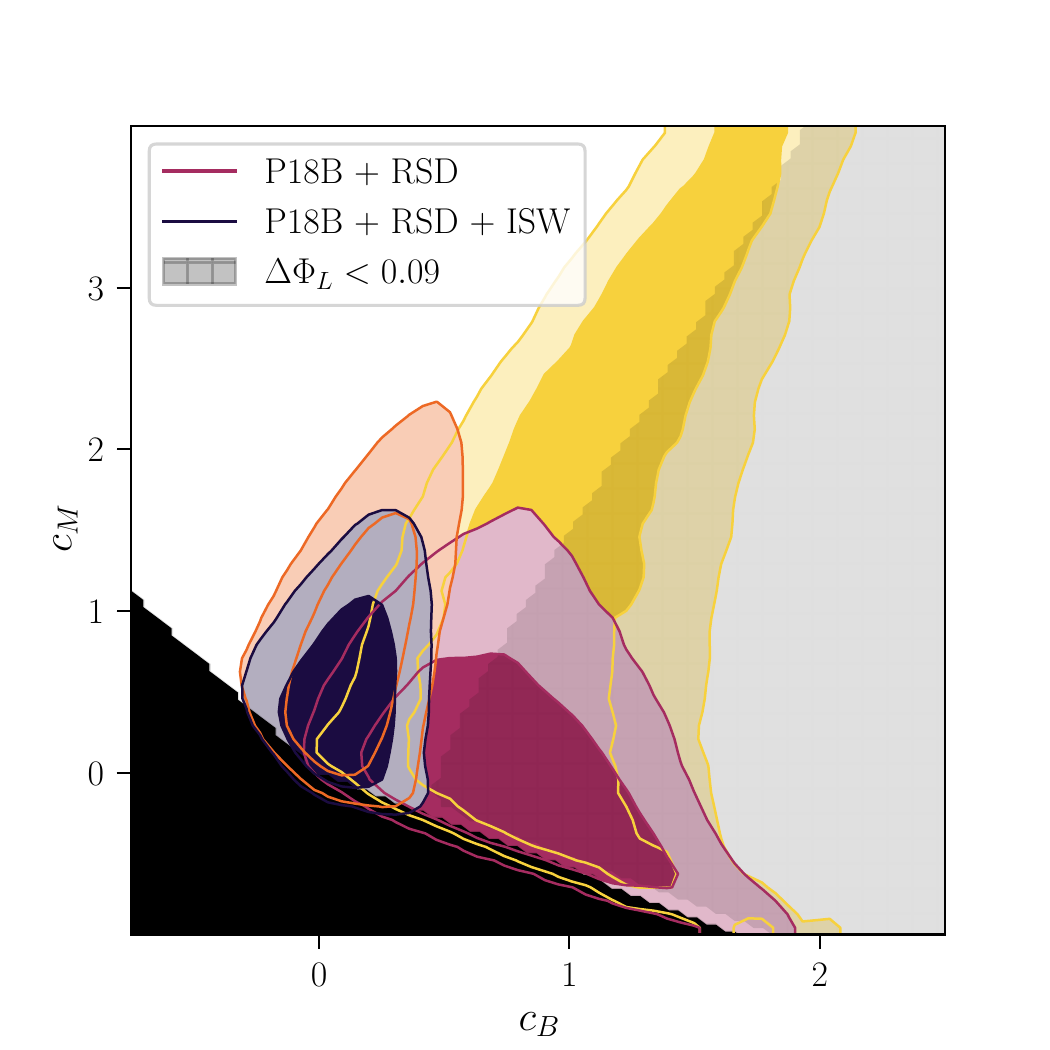}
    \caption{Analogous plot to figure \ref{fig-comboPlot}, but showing a different (more restrictive) ISW exclusion prior. Black regions are excluded by the presence of (gradient) instabilities, while the hashed out grey area corresponds to an ISW exclusion prior requiring that the lensing potential decays by at least $9\%$ between redshifts $5$ and $0$. In terms of the $\Delta\Phi$ notation introduced in equation \eqref{eq-DeltaPhi}, this means we require $\Delta\Phi_L > 0.09$, i.e. rule out regions of parameter space where $\Delta\Phi_L \leq 0.09$. Here we also show the effect of including RSD constraints explicitly. Noticeably, most of the P18B + RSD favoured parameter space is inconsistent with the ISW exclusion prior shown.}
    \label{fig-comboPlot-DeltaPrior}
\end{figure}

\section{ISW priors from observational constraints}
\label{app-ISWpriors}

In the main text we investigated both qualitative ISW constraints derived from the simple requirement that the lensing potential $\Phi_L$ is decaying as well as observational constraints derived by computing the precise (ISW) cross-correlations between galaxies and CMB temperature, cf. figures \ref{fig-ObservationalConstraints} and \ref{fig-comboPlot}. Here we now wish to investigate whether we can identify a more restrictive criterion/theory prior to demarcate the region of parameter space consistent with ISW constraints than the conservative bound discussed in the main text that simply mandated a decaying lensing potential $\Phi_L$. 
Ideally, such a criterion can recover the observational constraints more closely even before computing any galaxy-CMB temperature cross-correlations and hence serve as an effective theoretical prior. Such a prior would provide a powerful shortcut in the analysis of dark energy models in the future, so has the potential to be very helpful in practice.  

In figure \ref{fig-comboPlot} we showed P18B and P18B+ISW data constraints in the marginalised $\{c_b,c_m\}$ plane, combined with the qualitative `exclusion prior' from demanding a decaying lensing potential. It is clear that the P18B+ISW constraints are highly consistent with this prior, also providing a helpful consistency check for the ISW likelihood implementation. Let us formalise an exclusion prior like the one discussed above by defining $\Delta \Phi_L$ as follows 
\begin{align}
    \Delta \Phi_L \equiv \frac{\Phi_L{}|_{z=5}-\Phi_L{}|_{z=0}}{\Phi_L{}|_{z=5}},
    \label{eq-DeltaPhi}
\end{align}
where $X|_{z=z_0}$ denotes the evaluation of the function $X$ at redshift $z_0$. $\Delta \Phi_L$ in effect measures whether the lensing potential overall decays/grows between redshifts $5$ and $0$ and, if so, by how much. The qualitative ISW constraints discussed in section \ref{subsec-qualitativeISW} therefore amount to placing a $\Delta\Phi_L > 0$ prior (since a decay in time implies growth with increasing redshift). Note that the precise choice of the upper redshift chosen will only very mildly affect the analysis that follows. Conceptually speaking, what is important is to choose a redshift before dark energy plays a significant role in determining the Universe's expansion -- we choose redshift $5$ here as an illustrative example.   

\begin{figure*}[ht!]
    \centering
    \includegraphics[width=0.49\linewidth]{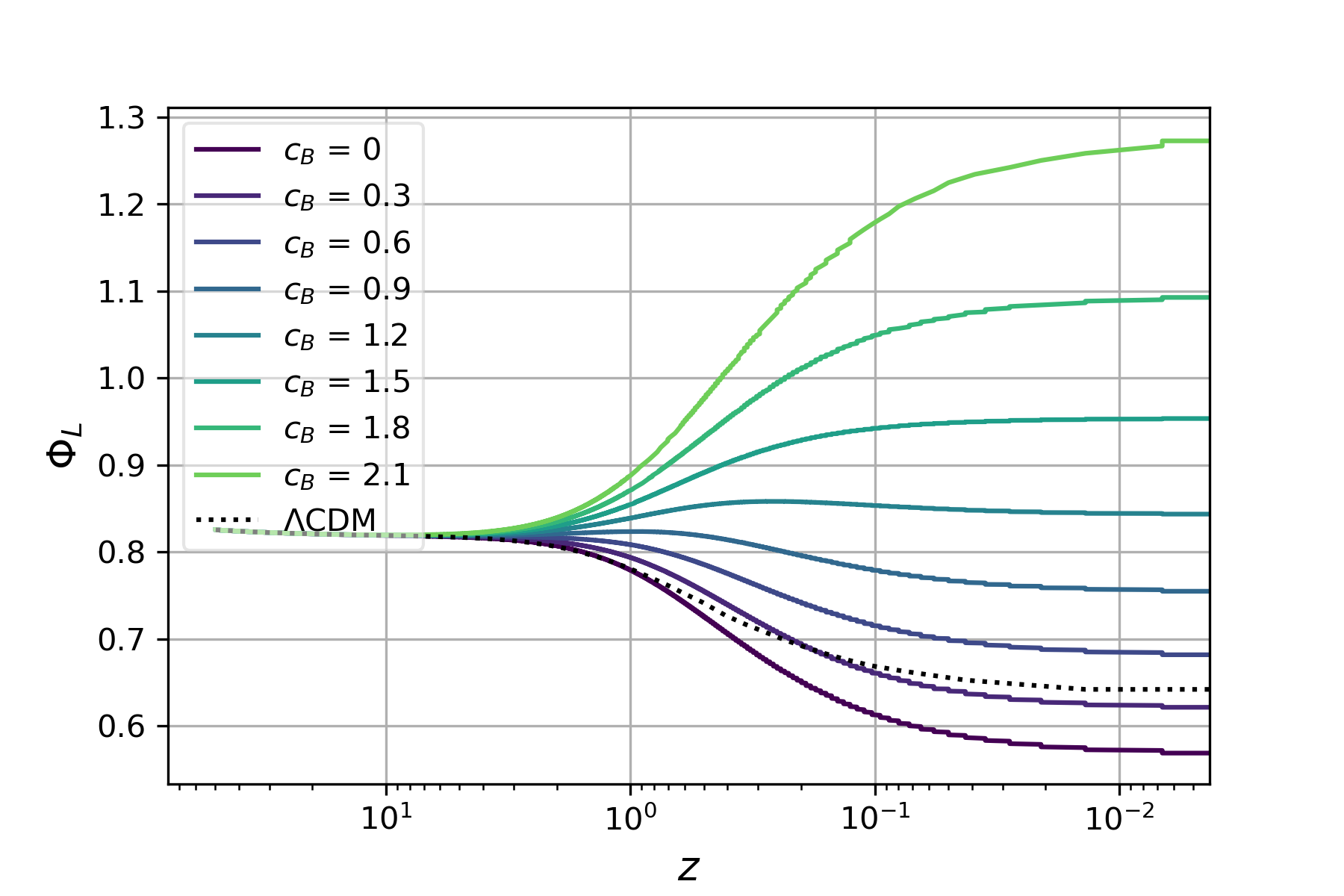}
     \includegraphics[width=0.49\linewidth]{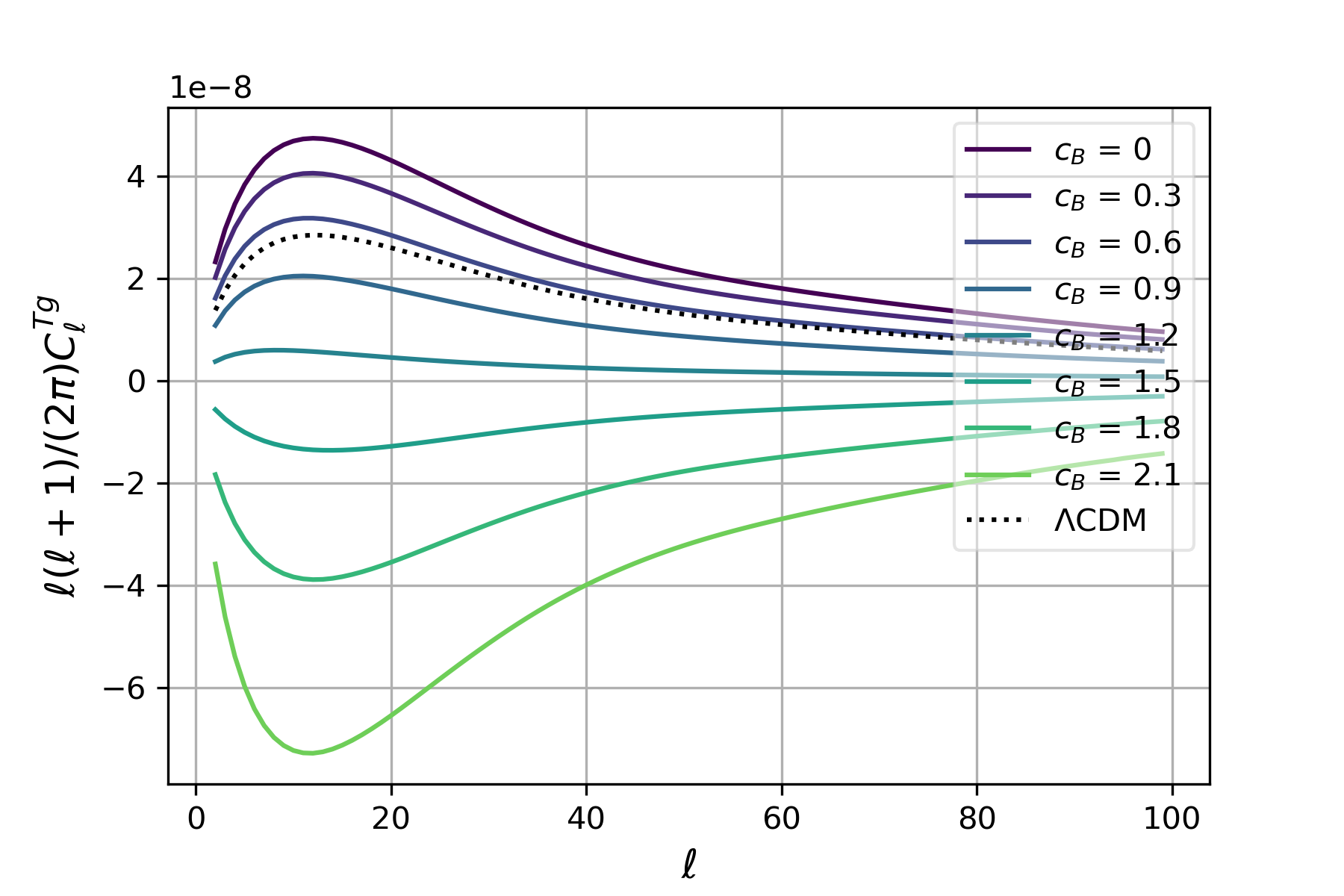}
    \caption{
    {\bf Left plot}: Evolution of the lensing potential $\Phi_L=\Phi+\Psi$ as a function of redshift for various choices of the dark energy parameter $c_b$, where $\alpha_i=c_i \Omega_{DE}$ -- see equation \eqref{alphadef}. All $\Lambda{}$CDM parameters are fixed to the P18 best-fit cosmology \cite{Planck:2018vyg} and in this plot we show results for a fiducial choice of the second dark energy parameter $c_m = 1$ (compare with figure \ref{fig-lensingPotential} where this was fixed to $c_m = 0$). Here we also explicitly see that $\Phi_L$ can decay faster than in $\Lambda{}$CDM. 
    {\bf Right plot}: Here we show the $C_\ell^{Tg}$ quantifying cross correlations between galaxies and CMB temperature measurements for the same cosmologies as on the left -- c.f. figure \ref{fig-Clevolution} in the main text.
    }
    \label{fig-Clevolution-v2}
\end{figure*}
Figure \ref{fig-comboPlot} in the main text clearly showed that a $\Delta\Phi > 0$ prior unsurprisingly and significantly underestimates the strength of current ISW constraints. In figure \ref{fig-comboPlot-DeltaPrior} we here show the analogous exclusion region for a $\Delta\Phi > 0.09$ prior, i.e. the effect of a prior that demands a decay of the lensing potential by at least $9\%$ between redshifts $5$ and $0$. Two observations then jump out: 1) This prior indeed successfully rules out more of the parameter space ultimately excluded by ISW constraints and we can see that cosmologies with less than a $9\%$ decay in the lensing potential are indeed ruled out at $2\sigma$ here, 2) ISW data constraints clearly do more than just place a constraint on $\Delta\Phi$. In particular one can see that for $c_b, c_m \gtrsim 1$ regions consistent with P18B constraints at $1\sigma$ are consistent with $\Delta\Phi > 0.09$, yet are ruled out by ISW data. We find that such cases are e.g. associated with cosmologies, where $\Phi_L$ does indeed decay at early and late times, but plateaus for intermediate redshifts. Such evolutions are qualitatively different to e.g. those shown in figure \ref{fig-Clevolution-v2} and produce a significantly worse fit to the observed $C_{\ell}$'s. 

Overall we therefore see how, while imperfect, the above $\Delta \Phi_L$ can help to understand which regions of parameter space are consistent with ISW constraints. We caution that the more restrictive prior discussed here has been derived in the context of the Horndeski scalar-tensor theories we have focused on here, with implicit assumptions on the form of the underlying functional freedom encoded in the parametrisations used. We leave a more global analysis for different theories and parametrisations to future work, but caution that a specific $\Delta \Phi_L > X$ prior for some value $X$ should not be blindly applied to other analyses without 
ideally first verifying its accuracy in this new context in a similar fashion to what has been shown here. Finally, note that we have clearly tuned the $\Delta \Phi_L > 0.09$ prior discussed above {\it a posteriori} given the data constraints shown e.g. in figures \ref{fig-comboPlot} and \ref{fig-comboPlot-DeltaPrior}, so this is directly impacted by the constraining power of current data and an analogous prior should be adjusted as future data become available.

\section{Different \texorpdfstring{$\alpha_i$}{[alphai]} parametrisations} \label{app_proptoa}
As discussed in the main text, numerous parametrisations exist for the $\alpha_i$ EFT functions encoding the dynamics of linear perturbations and we have presented constraints focused on the commonly used $\alpha_i = c_i \Omega_{\rm DE}$ parametrisation. To complement this, here we show constraints analogous to what is displayed and discussed in the main text in section \ref{sec-cosmo} for another parametrisation that is commonly used, namely for $\alpha_i = c_i \cdot a$. As in the main text we show constraints in the $\{c_m,c_b\}$ plane as well as when those constraints are mapped to the $\{\mu,\Sigma\}$ parameter space. 

\begin{figure*}[t!]
    \centering
    \includegraphics[width = 0.48\textwidth]{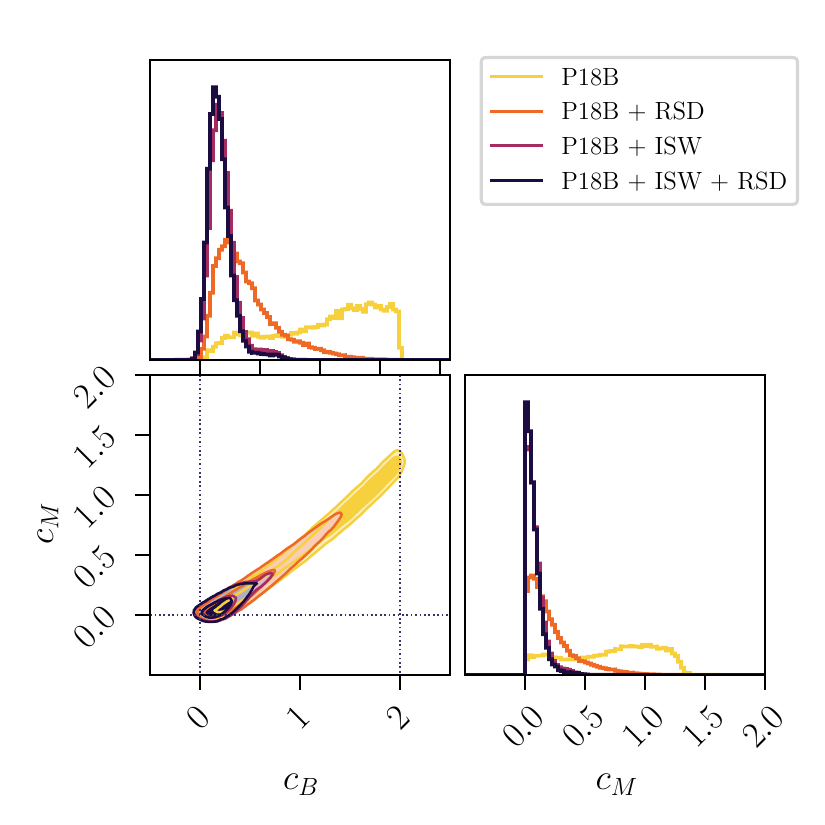}
    \includegraphics[width = 0.48\textwidth]{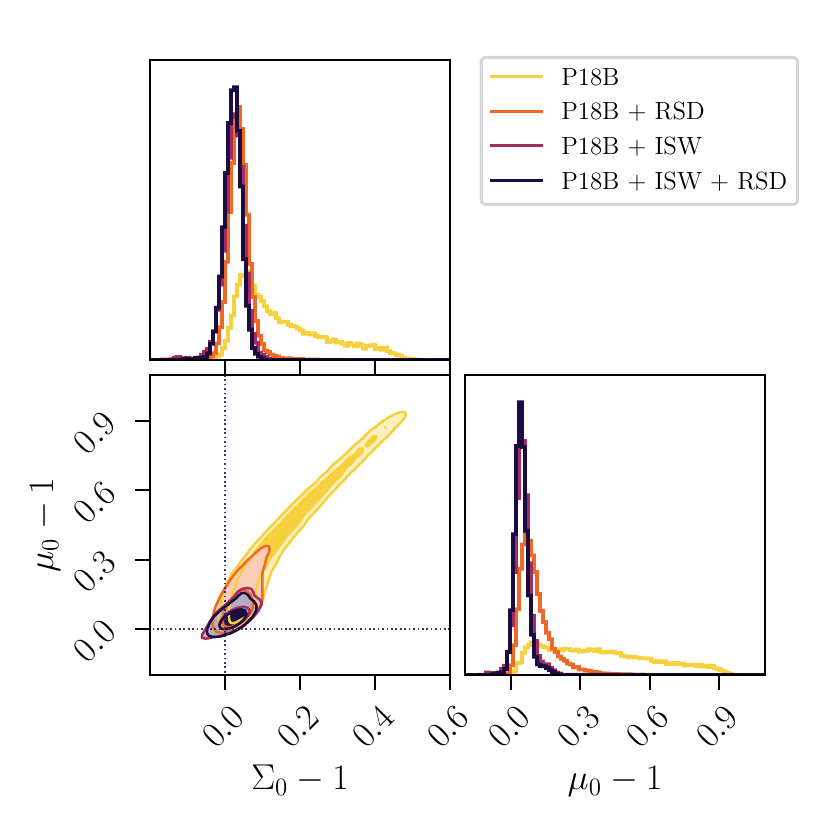}
    \caption{{\bf Left panel}: Here we show constraints in the $\{c_m,c_b\}$ plane for the $\alpha_i = c_i a$ parametrisation. So this is the analogue to the left panel of figure \ref{fig-ObservationalConstraints}, where this was shown for $\alpha_i = c_i \Omega_{\rm DE}$. In the parametrisation shown here, ISW constraints noticeably fully preempt any additional constraints placed by RSD measurements. Notice that we impose $\alpha_B \leq 2$ here -- see appendix \ref{app_proptoa} for a related discussion -- but this only affects the P18B-only bounds. 
    {\bf Right panel}: Analogous plot to the left panel, but for the more phenomenological $\mu/\Sigma$ parametrisation, where $\mu_0/\Sigma_0$ are related to the $c_i$/$\alpha_i$ via \eqref{muSigma}. Note that the $0$ subscript denotes the value of $\mu/\Sigma$ today.    
    }
    \label{fig-aParam}
\end{figure*}

Our main results for the $\alpha_i = c_i \cdot a$ parametrisation are shown in figure \ref{fig-aParam} and table \ref{tab_ci_proptoa_bounds}. Differences to constraints obtained using the $\alpha_i = c_i \Omega_{\rm DE}$ parametrisation in figure \ref{fig-ObservationalConstraints} and table \ref{tab_ci_bounds} are immediately apparent and neatly illustrate a point stressed in the main text: The choice of underlying parametrisation has a significant effect on the precise constraints obtained and one should therefore not expect identical constraints when using different parametrisations.\footnote{A more detailed investigation of this parametrisation dependence is beyond the scope of this paper, but we here simply report the corresponding bounds using the  $\alpha_i = c_i \cdot a$ for comparison and to illustrate this point.} 
For the $\alpha_i = c_i \Omega_{\rm DE}$ parametrisation we saw that the novel ISW constraints we focus on here (when combined with P18B bounds) were significantly stronger than the RSD constraints we compared them with. Nevertheless those two sets of constraints were complementary and P18B + RSD + ISW constraints were noticeably improved over P18B + ISW. For the  $\alpha_i = c_i \cdot a$ parametrisation this is not the case, with near identical bounds for both scenarios. for this parametrisation, the P18B + ISW constraints almost fully preempt any further improvements from adding RSD constraints in our setup. This is primarily due to the much tighter $\{c_m,c_b\}$ contours for this parametrisation (a consequence of the different time-dependence for the $\alpha_i$) and is mirrored when inspecting constraints in the $\{\mu,\Sigma\}$ plane, as shown in figure \ref{fig-aParam}. 

Finally, a quick note about prior boundaries. Already in the $\alpha_i = c_i \Omega_{\rm DE}$ parametrisation discussed in the main text, we implicitly encountered some hard prior boundaries,  where there is an excellent fit to the data on the boundary. For example in figure \ref{fig-comboPlot} gradient stability constraints place hard cuts in some directions in parameter space. Similarly, for the $\alpha_i = c_i \cdot a$ parametrisation considered in this appendix, $\alpha_M$ and hence $c_m$ cannot become negative due to gradient stability constraints and so while the data in fact prefer a near-zero value for $c_m$, stable deviations are only possible in the positive $c_m$ direction. In addition, we here encounter a second hard prior boundary: We have restricted $\alpha_B$ to remain smaller than 2 at all times. Crossing the $\alpha_B = 2$ point (which would happen at some point in the evolution for all models with $c_B \geq 2$) is associated with a number of discontinuities and so evolutions crossing this point have frequently been conservatively excluded in the literature \cite{Noller:2018wyv,SpurioMancini:2019rxy,Noller:2020afd}. Since this issue only affects the pure P18B constraints discussed in this appendix and becomes irrelevant once RSD or ISW constraints are added, we follow the same conservative methodology here and refer to \cite{Bellini:2014fua,Lagos:2017hdr,Ijjas:2017pei,Noller:2018wyv,SpurioMancini:2019rxy,Noller:2020afd} for more in-depth discussions of this point. 

\begin{table*}[t!] 
\renewcommand{\arraystretch}{1.8}
\setlength{\tabcolsep}{0.3cm}
\begin{tabular}{|l||c|c||c|c|}  \hline  
Data sets & $c_b$ & $c_m$ & $\mu_0 - 1$ & $\Sigma_0 -1$\\ \hline\hline
P18B & $0.29 < c_b < 2^{*}$ & $0^* < c_m < 1.23$ & $0.37^{+0.49}_{-0.32}$ & $0.15^{+0.28}_{-0.14}$\\ \hline
P18B + ISW & $0.23^{+0.42}_{-0.21}$ & $0*< c_m < 0.24$ & $0.05^{+0.09}_{-0.06}$ & $0.02^{+0.05}_{-0.05}$\\ \hline
P18B + RSD & $0.51^{+0.83}_{-0.42}$ & $0* < c_m < 0.68$ & $0.12^{+0.22}_{-0.10}$ & $0.04^{+0.07}_{-0.05}$\\ \hline
P18B + ISW + RSD & $0.21^{+0.39}_{-0.19}$ & $0* < c_m < 0.22$ & $0.05^{+0.09}_{-0.05}$ & $0.02^{+0.04}_{-0.05}$\\ \hline
\end{tabular}
\caption{
Here we show $2\sigma$ bounds (marginalised 1D posteriors) on the dark energy/modified gravity $c_i$ parameters in the $\alpha_i = c_i \cdot a$ parametrisation for different combinations of data sets, as discussed in appendix \ref{app_proptoa}. We also show the corresponding bounds on $\mu_0$ and $\Sigma_0$, where the $0$ subscript denotes the value of these parameters today and the parameters themselves are inferred from the $c_i$ (and ultimately $\alpha_i$) using the expressions discussed around equation \eqref{muSigma}.  When the $c_i$ distribution is strongly non-Gaussian and there is an excellent
fit to the data on a prior-induced boundary, we do not give a mean value and denote limit values 
with an asterisk -- see appendix \ref{app_proptoa} for details.}
\label{tab_ci_proptoa_bounds}
\end{table*}

\section{Cosmological parameter constraints beyond EFT parameters}\label{app-LargeComp}

\begin{figure*}[ht!]
    \centering
    \includegraphics[width=0.99\linewidth]{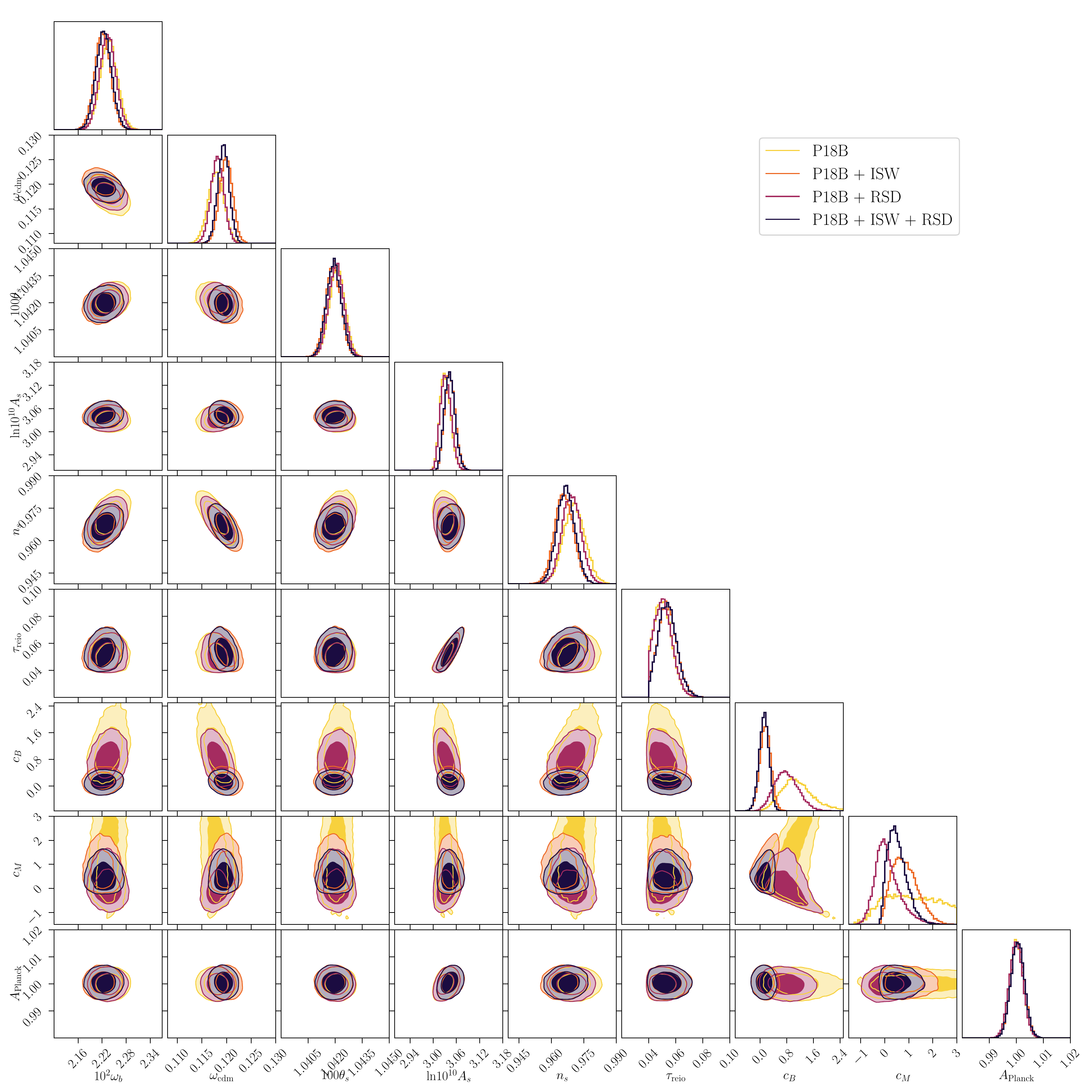}
    \caption{
    Here we show parameter constraints for the full dark energy $\{c_b,c_m\}$ + standard $\Lambda{}$CDM parameter space for the same combinations of data sets as in the main text. Contours mark $68\%$ and $95\%$ confidence intervals, and are computed using Planck 18 and BAO data (P18B), as well as measurements of galaxy-galaxy auto- and galaxy-CMB temperature cross correlations (ISW), and redshift space distortion measurements (RSD) -- for details on the data sets used see section \ref{sec-data}. See appendix \ref{app-LargeComp} for a discussion of the constraints shown.
    }
    \label{fig_oISW_CompLarge}
\end{figure*}

Throughout the above we have focused on constraints for the $c_i$ parameters controlling the $\alpha_i$ functions and the corresponding constraints when mapped to the $\{\mu,\Sigma\}$ plane, always marginalised over the standard cosmological $\Lambda{}$CDM parameters as well as any `nuisance' parameters associated with the bias modelling. In figure \ref{fig_oISW_CompLarge} we now for completeness also show constraints for the standard cosmological $\Lambda{}$CDM parameters for one of the parametrisations used, namely $\alpha_i = c_i \Omega_{DE}$. As pointed out in the main text, we see that the addition of ISW constraints only has a mild effect on the standard $\Lambda{}$CDM parameters, most noticeably a small shift in $\omega_{\rm cdm}$ as a result of larger $c_b$ values mildly preferring smaller $\omega_{\rm cdm}$. However, no strong degeneracies between the $c_i$ and standard $\Lambda{}$CDM parameters are present. For more details on how these constraints compare to pure $\Lambda{}$CDM constraints in the absence of the deviations encoded in the $c_i$, we refer to the analysis of \cite{Noller:2018wyv}. Finally we note that, motivated by observations of the Gunn-Peterson trough (see e.g. \cite{SDSS:2001tew}), in our analysis we have imposed a prior $\tau_{\rm reion} \geq 0.04$, corresponding to $z_{\rm reion} \gtrsim 6$.

\bibliographystyle{utphys}
\bibliography{EFTofDE_ISW}
\end{document}